\pdfoutput=1
\documentclass[10pt,journal]{IEEEtran}
\usepackage{graphicx}
\usepackage{booktabs}
\usepackage{amsmath}
\usepackage[hidelinks]{hyperref}
\graphicspath{{figs/}}

\DeclareFontShape{OT1}{ptm}{m}{scit}{<->ssub*ptm/m/it}{}
\DeclareFontShape{OT1}{ptm}{b}{scit}{<->ssub*ptm/b/it}{}
\IfFileExists{ts1ptm.fd}{\input{ts1ptm.fd}%
  \DeclareFontShape{TS1}{ptm}{m}{sc}{<->ssub*ptm/m/n}{}%
  \DeclareFontShape{TS1}{ptm}{b}{sc}{<->ssub*ptm/b/n}{}}{}

\begin{document}

\title{Edge-Inference Governors Need Memory-Clock State}

\author{Jaehoon~Kang%
\thanks{J.~Kang is with Ecorbit. E-mail: dk@ecorbit.com.}%
\thanks{This is the extended technical report: the supplementary
sections S1--S12 referenced throughout (``Supp.~\S S$n$'') are included
here as appendices, making this version self-contained.}}

\markboth{Kang: Edge-Inference Governors Need Memory-Clock State (extended technical report)}%
{Kang: Edge-Inference Governors Need Memory-Clock State (extended technical report)}

\maketitle

\begin{abstract}
On integrated edge SoCs whose memory fabric is governed independently of the compute clocks, frequency-aware latency estimators let deadline-aware DVFS governors schedule ML inference by modeling latency over CPU and GPU clocks --- but they do not condition on the memory clock (EMC), a deployment state omitted from their state vector that decides whether a governor meets its deadlines and at what energy. We show this with a deployed, measured governor on Jetson Orin: an EMC-blind GPU-only fit misses $25$--$28\%$ of cycles at tight deadlines, whereas an EMC-aware two-cell refit holds misses to ${\le}0.9\%$ under a $2\%$ QoS budget --- selecting a budget-feasible operating point \emph{proactively}, before the first release, where latency-only reactive calibration updates only from completed responses and cannot repair memory-clock-dependent slope error. Across six models deployed over two Orin SKUs, the core MobileNetV2 and ViT-Small results replicate on both boards, while detection and broader LLM deployments reproduce the failure on the NX. Sustained deployment then requires two further state layers: decode horizon --- KV-cache growth erodes tight-deadline feasibility over long responses --- and GPU co-tenancy, whose occupancy opens queueing tails that median-accurate models cannot express. A contract-admission policy composes the three layers from bounded probes and deploys live, including a joint cell decoding $2{,}000$ tokens against an active GPU co-tenant where tenancy-blind admission misses $32\%$ of tokens. Under the reference-probe maximum-guard accounting, every accepted contract is decisively measured-feasible. Fresh-probe stress tests then expose a probe-variance failure mode at knife-edge admissions; a dispersion-banded guard, validated on held-out trials, eliminates the observed probe-induced low-clock selections (held-out aggregate $1.19\%$, launch-level upper bound $1.70\%$); and one additional bin of headroom yielded observed per-launch compliance across eight fresh launches. A governor's guarantee is thus a measured conservatism ladder --- state repair, probe-variance banding, and headroom for launch realization. This complements, not rebuts, the state of the art within its CPU$\times$GPU scope.
\end{abstract}


\IEEEpeerreviewmaketitle

\section{Introduction}\label{sec:intro}

Deadline-constrained ML inference on embedded SoCs increasingly relies on
DVFS governors guided by \emph{frequency-aware latency estimators}: models
that predict inference latency across processor frequency settings and
pick the most efficient setting that meets the deadline. The state of
the art models each processor's independent time with a
frequency-dependent term of the form $T(f) = k/f + b$ --- where $b$
absorbs ``frequency-independent'' overheads, explicitly including
\emph{memory transfer delays} --- and combines those per-layer terms
with a model of asynchronous CPU--GPU interaction~\cite{flame2026}; its
modeled clock-state space is CPU$\times$GPU. Related estimators fit a
power law over GPU frequency alone~\cite{dvfsaware2025}. On integrated SoCs, however, the
memory subsystem has its own DVFS domain (the EMC on NVIDIA
Jetson~\cite{nvidiajetsondocs}), so $b$ is in fact $b(f_{\mathrm{emc}})$.
Our central claim is that the EMC clock is a \emph{missing deployment
state}: a variable absent from a CPU$\times$GPU estimator's state vector,
yet one that decides whether a deadline governor meets its deadlines and
at what energy. We do not claim the state of the art is incorrect within
its CPU$\times$GPU scope; we show deployed deadline control on integrated
SoCs depends on state that a CPU$\times$GPU estimator omits.

We make this concrete by \emph{deploying} the failure and its repair. On
an Orin NX, at tight deadlines near the workload floor, a GPU-only fit
blind to the EMC misses $25$--$28\%$ of cycles (three repeats), while the
EMC-aware refit --- two profiling cells at the deployment memory point ---
holds misses to ${\le}0.9\%$ at the lowest budget-feasible clock; across
six models, the core MobileNetV2 and ViT-Small deployments replicate on
both SKUs, and the detection and LLM deployments reproduce the failure
on the NX (\S\ref{sec:solo}). Sustained deployment then exposes two further state
layers the point model does not carry (\S\ref{sec:layers}): the
\emph{decode horizon} --- KV-cache growth drifts per-token latency so that
a clock feasible for a short response violates its budget late in a
$2{,}000$-token one --- and \emph{GPU co-tenancy}, which barely moves the
median but opens a queueing tail that misses $10$--$32\%$ of deadlines at
clocks every median-fit policy admits.

The deployed evaluation demonstrates that these three state categories
suffice to repair the large systematic admission failures in the
evaluated deployments; at knife-edge contracts, the desired guarantee
level then determines the required conservatism --- aggregate
feasibility at the banded pick, or observed per-launch compliance with
one additional bin of headroom (\S\ref{sec:account}). The vehicle is a
\emph{contract-admission} policy (\S\ref{sec:policy}): each layer
is restored by a bounded probe --- a two-cell EMC refit, a measured drift
coefficient, and a $200$-cycle occupancy probe --- composed into one
feasibility rule with a maximum-based probe guard, hardened by a
dispersion band after end-to-end stress testing
(\S\ref{sec:account}). Run live on the hardware, including a
joint-stress cell that decodes $2{,}000$ tokens against an active GPU
co-tenant, under the reference-probe maximum-guard accounting the policy's
acceptances all measure decisively feasible, its refusals err
conservative, and its stress tests expose --- and a held-out-validated
dispersion band eliminates --- its own probe-induced low-clock
selections (\S\ref{sec:account}).

\textbf{Contributions.}
(1)~\textbf{A deployed missing-state failure and its bounded repair.} An
EMC-blind GPU-only fit misses $25$--$28\%$ of cycles at tight NX deadlines
(and $100\%$ on the Nano at $D{=}9$\,ms), while an EMC-aware refit holds
misses to ${\le}0.9\%$ --- measured on the hardware with calibrated
$1$\,ms-sampled energy across six models spanning CNN, detection,
transformer, and LLM token decode --- MobileNetV2 and ViT-Small
replicated on both SKUs, the rest deployed on the NX --- and persistent
under a fused TensorRT fp16 engine (penalties roughly halve but persist at
$+8$--$21\%$). The repair is tabular and bounded: the EMC axis has four lockable
strata, and two-cell refits suffice for three of four characterization
workloads (the bandwidth-bound proxy needs a full per-point sweep) ---
a parametric frequency term is \emph{worse than no repair} for three of
four (\S\ref{sec:solo}).
(2)~\textbf{Two further measured deployment-state layers.} Decode horizon:
per-token latency drifts $+2.5$--$2.7\%$ over $2{,}000$ tokens,
multiplying the late-window miss rate from $0.83\%$ to $9.9\%$ at a
deadline the short-window profile accepts. Co-tenancy: a 10\,Hz vision co-runner moves
the victim median by little while $31\%$ of decode tokens collide onto a
$42$--$43$\,ms plateau (\S\ref{sec:layers}).
(3)~\textbf{A guarded contract-admission deployment.} Composing the three
state layers into one deployed contract-admission policy --- per-EMC-point
model, measured drift coefficient, clock-scaled occupancy margin, each
from a bounded probe --- makes every guarded-policy acceptance decisively measured-feasible
across the seven evaluated contracts, while a point-estimate variant
separately accepts one knife-edge vision cell that remains statistically
unresolved at the budget boundary --- including a joint-stress cell decoding $2{,}000$ tokens
against an active GPU co-tenant ($0.11\%$ measured at its first accepted
deadline, nine independent runs); for each omitted state category, at
least one evaluated contract causes the corresponding ablation to
violate the budget by $20$--$32\%$. End-to-end stress tests with a held-out-validated
dispersion band then decompose admission uncertainty into three
measured layers: omitted state ($20$--$32\%$), probe-induced low-clock
admission ($3\%$ at the knife-edge; eliminated by the band in held-out
trials), and condition realization ($0$--$2.9\%$ per launch around a $1.19\%$
held-out aggregate; one additional bin of headroom yielded observed
per-launch compliance in eight fresh launches)
(\S\ref{sec:policy}--\S\ref{sec:account}).

\begin{figure*}[t]\centering
\includegraphics[width=\textwidth]{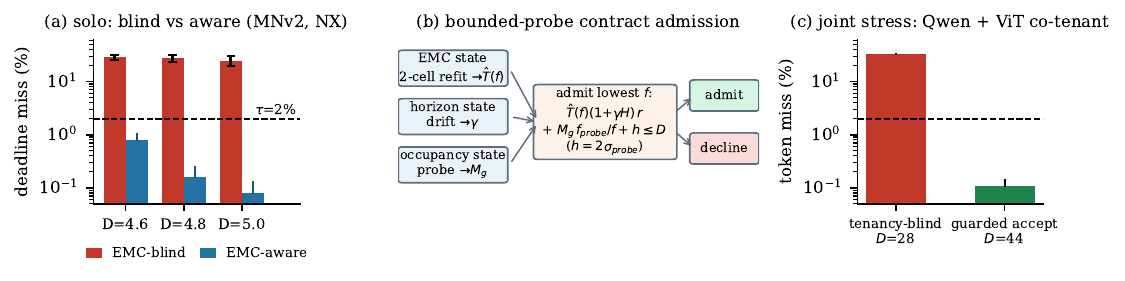}
\caption{\textbf{State omission, policy failure, and guarded admission.}
\textbf{(a)}~Solo condition (NX, MobileNetV2, EMC pinned at the
2133\,MHz deployment point): the EMC-blind GPU-only fit selects clocks
that miss $25$--$28\%$ of deadlines; the EMC-aware two-cell refit holds
${\le}0.9\%$, below the $2\%$ budget (dashed).
\textbf{(b)}~The composed contract-admission rule: each state layer is
restored by a bounded probe --- a two-cell EMC refit ($\hat{T}(f)$), a
measured decode-drift coefficient ($\gamma$), and a $200$-cycle occupancy
probe ($M$, clock-scaled by $f_{\mathrm{probe}}/f$, guarded by the
probe's sample maximum plus a dispersion band, \S\ref{sec:account};
$r$ is the fixed solo safety multiplier of \S\ref{sec:solo}) --- and a contract $(D,H)$ is admitted at the lowest
feasible clock or declined.
\textbf{(c)}~Joint stress (Qwen2.5-1.5B decoding $2{,}000$ tokens against
a live 10\,Hz ViT co-tenant): tenancy-blind admission at $D{=}28$\,ms
misses $32.3\%$ of tokens; the composed rule's first accepted deadline
($D{=}44$\,ms, 1173\,MHz) measures $0.11\%$ (run-level $95\%$ upper
bound $0.14\%$ over nine independent deployments).}\label{fig:overview}
\end{figure*}

\section{Background and Related Work}\label{sec:related}

\textbf{Frequency-aware latency estimation.}
FLAME~\cite{flame2026} exemplifies the state of the art: each
processor's independent per-layer time is modeled as $T(f) = k/f + b$,
where $b$ captures frequency-independent overheads --- in the authors'
words, ``pipeline stalls and cache misses on CPU, or kernel launch
latencies and memory transfer delays on GPU'' --- and these terms are
combined with a model of asynchronous CPU--GPU coupling. The modeled
frequency space is CPU$\times$GPU; the memory clock is not an input
dimension. Within that scope the estimator is
accurate and its deadline-aware governor improves on learning-based
adaptation in the zTT/ALERT lineage~\cite{flame2026,ztt2021,alert2020};
earlier
work models inference latency over GPU frequency
alone~\cite{dvfsaware2025}. Evaluation in this line is by
aggregate QoS percentages, and online adaptation is latency-level drift
calibration --- an EWMA-smoothed mean-level correction, not a model of
the distribution around the estimate. Memory-frequency scaling as an
independent power knob dates to server-class memory
DVFS~\cite{memscale2011}; joint memory--compute frequency scaling is
known to matter for \emph{energy} on Jetson~\cite{jointmemfreq2025},
and device-parameter sweeps there have measured the energy impact of
CPU/GPU and memory settings~\cite{stonybrookjetson} without modeling
latency over the memory axis or analyzing tails. What has been missing
is what omitting the axis does to \emph{latency estimation and deadline
guarantees}, and what a governor must carry in its state vector to
avoid the resulting failures.

\textbf{Admission control and QoS contracts.}
Treating deployment as per-contract admission has a long real-time
lineage: measurement-based admission control accepts or declines flows
from observed behavior rather than declared worst
cases~\cite{jamin1997}; constant-bandwidth servers and resource
kernels~\cite{abeni1998,rajkumar1998} reserve capacity so an admitted
task's contract survives co-tenants; and elastic task
models~\cite{buttazzo2002} negotiate rates when a contract cannot be
met. SLO-aware inference serving brings the same shape to ML systems:
Clockwork~\cite{clockwork2020} uses predictable execution-time
estimates to schedule requests toward strict request-level SLOs, and
later serving systems preserve SLOs under unpredictable load via
preemption and batching~\cite{shepherd2023}. Our policy is measurement-based admission for
on-device inference with the platform's own clock state in the loop: the
admission test is built from bounded probes of the deployed condition
(memory point, decode horizon, occupancy), and its guard is a maximum-based empirical guard, motivated by
order-statistic tolerance-limit reasoning~\cite{wilks1941} but not
claimed as a calibrated tolerance bound under correlated collisions. On-device autoregressive decode is widely memory- and
bandwidth-intensive~\cite{decodemembound,decodemembound2,melting2024,llmnpu2025};
we therefore use it as the memory-sensitive joint-stress case.

\textbf{Real-time GPU inference and interference.}
DARIS~\cite{daris}, DeepRT~\cite{deeprt}, and RTGPU~\cite{rtgpu}
report per-policy miss rates or schedulability for concurrent GPU tasks
and DNN inference, and kernel-level preemption systems such as
REEF~\cite{reef2022} protect latency-critical inference from
best-effort co-tenants. On integrated CPU--GPU SoCs,
memory-bandwidth-intensive co-runners slow GPU kernels by up to
$3\times$~\cite{gpumemcontention}; our contention design follows this
lineage, using IsolBench-style bandwidth
adversaries~\cite{isolbench}, and recent profiling of concurrent vision
inference on Jetson reports DVFS-driven anomalies under
co-tenancy~\cite{orinprofiling2025}. Bandwidth
reservation~\cite{memguard} mitigates such interference; our governor
instead treats the memory clock itself as admission state. These works report aggregate miss rates, yet
$(m,k)$-firm and weakly-hard analysis establishes that miss
\emph{patterns}, not rates, determine whether a control loop
survives~\cite{hamdaoui1995,bernat2001,maggio2020} --- one reason our
statistics use independent deployment runs and run-level bounds
throughout (\S\ref{sec:setup}).

\textbf{Tail-aware reporting.}
Tail-first latency reporting~\cite{percentilereporting} and MLPerf's
measurement discipline~\cite{mlperfinference} are standard, and
measurement-based probabilistic timing analysis fits extreme-value models
to observed execution times with well-surveyed caveats about
dependence~\cite{davissurvey}. Percentiles, however, summarize the
marginal distribution; our occupancy margins are validated per deployed
contract, and the guard avoids distributional fits entirely.

\section{System Model and Setup}\label{sec:setup}

\begin{table}[t]
  \centering
  \caption{Platforms and primary workloads. p50 is the median compute
  latency at EMC 3199\,MHz with CPU/GPU pinned (Nano; 1k iterations).
  Breadth deployments add ResNet-50, YOLOv8s, and Qwen2.5-3B
  (\S\ref{sec:solo}). Models and
  runtimes:~\cite{mobilenetv2,vit,resnet,yolov8,qwen25,onnxruntimesw,llamacppsw};
  AI $=$ realized arithmetic intensity~\cite{roofline}.}
  \label{tab:setup}
  \footnotesize
  \setlength{\tabcolsep}{4pt}
  \resizebox{\columnwidth}{!}{%
  \begin{tabular}{@{}ll@{}}
    \toprule
    \multicolumn{2}{@{}l}{\textbf{Platforms} (both: four lockable EMC points $\{204, 665.6, 2133, 3199\}$\,MHz)}\\
    Orin Nano Super & 8\,GB LPDDR5, GPU ${\le}1020$\,MHz, L4T R36.5\\
    Orin NX & 16\,GB LPDDR5, GPU ${\le}1173$\,MHz, L4T R36.4.3\\
    \midrule
    \textbf{Workload} & \textbf{runtime / character / p50}\\
    MobileNetV2  & ONNX Runtime CUDA fp32 / mixed / 4.20\,ms\\
    ViT-Small    & ONNX Runtime CUDA fp32 / mixed / 11.40\,ms\\
    Qwen2.5-1.5B Q4 & llama.cpp CUDA / bandwidth-bound decode / 24.26\,ms/tok\\
    GEMV decode proxy & ORT fp16 / bandwidth-bound / 4.54\,ms\\
    GEMM (L2-resident) & ORT fp16 / compute-bound / 18.60\,ms\\
    GEMM ($4{\times}2048^3$) & ORT fp16 / mixed (realized AI ${\sim}65$) / 10.96\,ms\\
    \bottomrule
  \end{tabular}}
\end{table}

We measure on two NVIDIA Jetson Orin SKUs (Table~\ref{tab:setup}) that
expose the \emph{same} four lockable EMC points: the requested and
verified EMC operating points are matched across SKUs, while
surrounding platform differences (GPU ceiling, L4T release, carrier)
may still affect the magnitude and qualitative manifestation of a
matched EMC step. CPU
cores are pinned via cpufreq and the GPU via devfreq, leaving the EMC ---
the memory controller's own DVFS domain --- as the manipulated variable.
Clock control fails silently on the tested BSPs, so every cell is
gated: on them, EMC locks held only with \texttt{bwmgr\_halt} asserted
alongside \texttt{mrq\_rate\_locked}, and off-table requests round
\emph{up} silently while the write appears to succeed; our launcher locks and reads
back every rate before the first cell and verifies each run against the
tegrastats clock trace (the full pitfall catalogue is Supp.~\S S1). The four-point lockable set also bounds the repair's cost:
adding the EMC axis multiplies profiling by four, not by a continuum.
The deployment scenario is the platform's own default behavior, not an
adversarial setup: the 15\,W and 25\,W nvpmodel power profiles~\cite{nvidiajetsondocs}
pin the EMC at 2133 and 3199\,MHz respectively, with zero EMC transitions
observed under load --- so the same binary, profiled under one profile
and deployed under the other, crosses memory points without any input a
CPU$\times$GPU estimator can see.

Each deployment cell runs a periodic-release loop: a \texttt{SCHED\_FIFO}
thread, pinned and memory-locked, releases one inference per period at
absolute times ($t_0 + iP$), so an overrun never delays subsequent
releases and queueing tails under co-tenancy are GPU behavior, not
harness backlog. Full per-cycle distributions are recorded, so a deadline
is a post-hoc analysis parameter; a thermal gate holds each cell below
$55\,^\circ$C at start, and a one-hour saturated-decode soak shows the
once-picked operating point stays feasible across a
${>}13\,^\circ$C excursion (every $2{,}000$-token window at
$0.79$--$0.90\%$, no throttling). Energy integrates the onboard INA3221
\texttt{VDD\_IN} module rail at its $1$\,ms update interval into
calibrated per-inference joules.
Throughout, the unit of replication is the independent deployment run ---
within-run cycles are serially correlated (clustering: \S\ref{sec:solo}), so
run-level repeats, not within-run binomial intervals, are the honest error
bar; $\pm$ denotes the s.d.\ (or, where labeled, min--max) across such runs,
and for zero-event cells we report the observed zero misses across
the stated independent runs --- we do not treat pooled correlated
cycles as a calibrated binomial confidence bound.
The NX headline cells carry thirteen runs --- ten confirmatory
deployments accumulated after the initial three-run comparison, all
retained. Commanded clocks take effect with an actuation lag of roughly
$1/5/8$\,ms (CPU/GPU/EMC) --- longer than every deadline deployed here
--- so operating points are selected \emph{before} a contract's first
release rather than repaired within a cycle.

We use ``governor'' in the deployment
sense: a policy that selects the operating point for a deadline-constrained
run, not a per-inference online controller --- actuation lag makes \emph{same-cycle} EMC rescue impractical at these
deadlines, so online adaptation must actuate one release ahead (we deploy
and measure both an online calibrator and a live EMC switch, \S\ref{sec:account}). We likewise use
``feasible'' throughout this section in the QoS-governor sense: an operating
point is feasible if its measured deadline-miss rate stays below the
deployment miss budget $\tau$ (we use $\tau{=}2\%$, the threshold the deployed
governor enforces), not hard-real-time zero-miss feasibility --- in serving
terms, a per-inference latency SLO with a $98\%$ attainment target, here for
single-stream on-device inference. We use
$\tau{=}2\%$ only as the deployed QoS budget; the blind-vs-aware separation
is insensitive to it --- the blind policy misses by tens to hundreds of
percent while the aware policy stays near zero --- though a stricter budget
would demand a larger safety margin on the tightest cell. The primary contribution is correspondingly not a
more elaborate optimizer: it is identifying the missing deployment state
--- a widely assumed CPU$\times$GPU abstraction omits it --- and
demonstrating, through a simple bounded-probe contract-admission policy,
that restoring that state changes deployed feasibility.

\section{The Solo Failure and Its Repair}\label{sec:solo}

\begin{figure}[t]\centering\includegraphics[width=\linewidth]{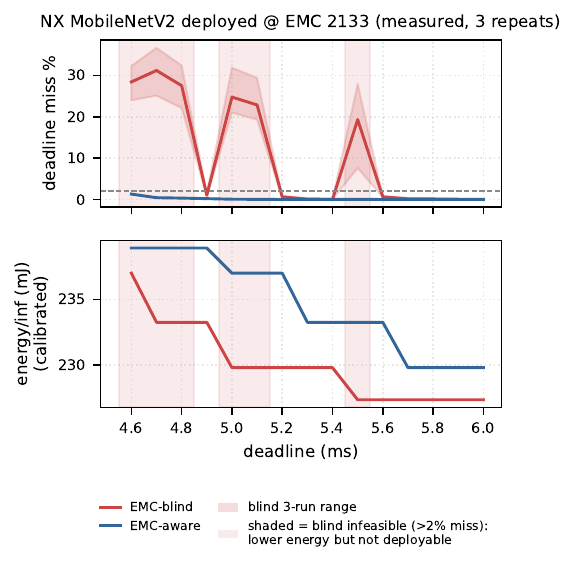}
\caption{Deployed EMC-aware vs.\ EMC-blind governor (NX, MobileNetV2, EMC
pinned at 2133\,MHz), measured over three independent $1{,}000$-cycle
deployments. Near the workload floor the blind policy (profiled at
3199\,MHz) misses $25$--$28\%$ (shaded band $=$ 3-run range); the
EMC-aware refit holds ${\le}0.9\%$ at the lowest budget-feasible
clock.}\label{fig:govdeploy}\end{figure}

A deadline-constrained deployment must pick a GPU clock that still meets
the deadline at whatever memory clock the power profile pins --- and that
is exactly the choice a CPU$\times$GPU estimator gets wrong. Each policy
is a (point model, margin) pair: given deadline $D$, select the lowest
GPU frequency with $T(f) + M \le D$, $T(f) = k/f + b$, $M = T(r{-}1)$,
where $r{=}1.035$ is the fixed solo-condition safety multiplier,
workload-independent, shared by every policy in this paper, fixed
\emph{before} any policy evaluation in this study, and held constant
during contract admission. It upper-bounds the largest solo locked-clock
p99/p50 dispersion we observe at the profiling point (${\le}1.9\%$)
with roughly $2\times$ headroom, and admission outcomes are insensitive
to it: replaying every evaluated contract at
$r \in \{1.02, 1.035, 1.05\}$ changes no co-tenancy pick and produces
no false admission --- a larger $r$ only trades utilization
(Supp.~\S S2).
The \emph{blind} policy carries $(k,b)$ fit at EMC 3199\,MHz, as an
estimator profiled in the 25\,W mode would; the \emph{aware} policy
carries a two-cell refit at the deployment point 2133\,MHz. Both then run
on the device at the clock each chose, and we measure misses on the wire
and joules on the rail.

The blind fit under-predicts the deployed latency and steps below the
miss cliff (Fig.~\ref{fig:govdeploy}, Table~\ref{tab:govdeploy}):
$28.4{\pm}3.4\%$ of cycles missed at $D{=}4.6$\,ms, $27.5{\pm}4.2\%$ at
$4.8$\,ms, $24.8{\pm}5.0\%$ at $5.0$\,ms, against ${\le}0.9\%$ for the
aware refit (run-level $95\%$ upper bounds ${\le}1.04\%$); on the Nano
the blind model's few-hundred-microsecond optimism selects a clock that
misses $100\%$ of the time while the aware policy misses $0.1\%$.

\begin{table}[t]
  \centering
  \caption{Deployed MobileNetV2 governor at the 15\,W memory point (EMC
  2133\,MHz). The \emph{blind} policy is fit at EMC 3199\,MHz; the
  \emph{aware} policy is a two-cell refit at 2133\,MHz. $f^\star$ is the
  selected GPU clock. Miss rates are measured over independent $1{,}000$-cycle
  deployments (the unit of replication); blind $=$ mean${\pm}$s.d.\ over 3
  runs, aware $=$ mean over 13 runs (NX) or 3 runs (Nano) with the run-level
  bootstrap~\cite{efron1979} 95\% upper bound in brackets --- all below the $\tau{=}2\%$
  budget. Energy is calibrated $1$\,ms-sampled \texttt{VDD\_IN} ($n{=}2$
  repeats per NX cell, $n{=}3$ Nano; sign confirmed in 10/10 interleaved
  A/B pairs).}
  \label{tab:govdeploy}
  \footnotesize
  \resizebox{\columnwidth}{!}{%
  \begin{tabular}{@{}llrrrr@{}}
    \toprule
    & & \multicolumn{2}{c}{blind (fit @3199)} &
        \multicolumn{2}{c}{aware (2-cell @2133)} \\
    \cmidrule(lr){3-4}\cmidrule(lr){5-6}
    SKU & $D$ & $f^\star$ & miss\% & $f^\star$ & miss\% [95\% UB] \\
    \midrule
    NX   & 4.6\,ms & 1122\,MHz & $28.4{\pm}3.4$ & 1173\,MHz & $0.81$ [$1.04$] \\
    NX   & 4.8\,ms & 1020\,MHz & $27.5{\pm}4.2$ & 1173\,MHz & $0.16$ [$0.25$] \\
    NX   & 5.0\,ms & 918\,MHz & $24.8{\pm}5.0$ & 1122\,MHz & $0.08$ [$0.13$] \\
    Nano & 9\,ms & 408\,MHz & $100.0$ ($\times3$) & 510\,MHz & $0.1$ [$0.3$ worst run] \\
    \midrule
    \multicolumn{2}{@{}l}{$\Delta$ energy (aware$-$blind)}
      & \multicolumn{4}{r}{NX $+1.9$ to $+7.2$; Nano $+1.2$\,mJ/inf (aware higher; low-energy feasible)} \\
    \bottomrule
  \end{tabular}}
\end{table} The failure is not
graceful: locked-clock response distributions are knife-edge
(p99.99/p50 ${\le}1.10$ in all eight 100k-cycle characterization cells),
so a few hundred microseconds of estimation bias separate near-full QoS
from tens-of-percent misses. Misses also arrive \emph{clustered} ---
continuation probability up to $0.74$ against $0.001$ under independence
($360$--$740\times$, bursts to 16 consecutive) --- which is why every
number in this paper is replicated at run level rather than by within-run
intervals, and why the guard of \S\ref{sec:policy} uses an
order statistic rather than a resampling estimate.

Nor is the profile pair special. Across all six directed profile pairs
over $\{665.6, 2133, 3199\}$\,MHz, the failure splits by direction:
fitting at a higher memory clock than deployed under-predicts and picks
infeasible clocks ($24.8\%$ miss into 2133\,MHz; $100\%$ into 665.6,
where the aware refit still holds $0.5\%$); fitting at a lower clock
over-predicts and over-provisions --- the blind policy selects the
maximum clock where the aware refit shows a cheaper feasible one, turning
the same blindness into pure energy waste (all six pairs:
Supp.~\S S2). Both halves of the thesis ---
whether a governor meets its deadlines \emph{and at what energy} ---
appear as the two directions of one omission.

\begin{table}[t]
  \centering
  \caption{\textbf{Top:} deployed blind-vs-aware governor across six
  models and two SKUs (representative tight deployment cells per model --- a compact range
  summarizes adjacent tight cells where shown; measured miss\% over
  independent runs; aware column shows [run-level
  95\% UB] where $n{\ge}3$; $\Delta E$ = aware$-$blind calibrated energy).
  Across all six directed (profile, deployment) EMC pairs: 3/3 fit-high
  pairs miss $24.8$--$100\%$; 3/3 fit-low pairs over-provision
  (supplement). Nano companions: ViT-Small blind $34$--$100\%$, aware
  ${\le}0.07\%$. \textbf{Bottom:} held-out median/max p50 error (\%) at
  EMC 2133\,MHz --- GPU-only fit at 3199 (A), parametric
  ${+}m/f_{\mathrm{emc}}$ (B), two-cell refit at the deployment point,
  scored on non-training cells (C).}
  \label{tab:solo}
  \footnotesize
  \setlength{\tabcolsep}{3.5pt}
  \resizebox{\columnwidth}{!}{%
  \begin{tabular}{@{}llrrrrr@{}}
    \toprule
    & & \multicolumn{2}{c}{blind} & \multicolumn{2}{c}{aware} & \\
    \cmidrule(lr){3-4}\cmidrule(lr){5-6}
    workload (SKU) & $D$ & $f^\star$ & miss\% & $f^\star$ & miss\% [UB] & $\Delta E$\\
    \midrule
    MobileNetV2 (NX) & 4.8\,ms & 1020 & $27.5{\pm}4.2$ & 1173 & 0.16 [0.25] & $+7$\,mJ\\
    MobileNetV2 (Nano) & 9\,ms & 408 & $100$ & 510 & 0.1 [0.3] & $+1.2$\,mJ\\
    ViT-Small (NX) & 13.6--15\,ms & --- & 28--100 & --- & ${\le}0.2$ & $+5$--$13$\,mJ\\
    ResNet-50 (NX) & 14\,ms & 918 & $95.8{\pm}1.1$ & 1020 & 0.07 & $+11.8$\,mJ\\
    YOLOv8s (NX) & 30\,ms & 918 & $86.3{\pm}1.5$ & 1020 & 0.00 & $+24$\,mJ\\
    Qwen-1.5B (NX) & 29\,ms/tok & 918 & $100$ & 1020 & 0.8 [0.8] & $-10$\,mJ/tok\\
    Qwen-3B (NX) & 49.3\,ms/tok & 918 & $100$ & 1020 & 0.67 & $-19.5$\,mJ/tok\\
    \midrule
    \multicolumn{7}{@{}l}{\textbf{Held-out estimator error} (median / max \%, EMC 2133\,MHz)}\\
    & & \multicolumn{2}{r}{A: GPU-only} & \multicolumn{2}{r}{B: ${+}m/f_{\mathrm{emc}}$} & C: 2-cell\\
    \midrule
    MobileNetV2 & & \multicolumn{2}{r}{5.3 / 12.8} & \multicolumn{2}{r}{5.4 / 7.1} & 2.9 / 3.4\\
    ViT-Small & & \multicolumn{2}{r}{5.1 / 17.6} & \multicolumn{2}{r}{10.5 / 14.1} & 4.6 / 6.7\\
    decode proxy & & \multicolumn{2}{r}{3.4 / 32.2} & \multicolumn{2}{r}{19.9 / 30.9} & 12.7 / 20.9\\
    GEMM L2-res. & & \multicolumn{2}{r}{1.5 / 2.8} & \multicolumn{2}{r}{17.8 / 34.5} & 2.9 / 4.8\\
    \bottomrule
  \end{tabular}}
\end{table}

The failure is not a property of the models chosen
(Table~\ref{tab:solo}, top). Repeating the protocol on ResNet-50 --- a
workload the state-of-the-art estimator~\cite{flame2026} itself
evaluates on --- YOLOv8s,
and Qwen2.5 at 1.5B and 3B, the blind policy's selected clock has a
deployed median \emph{above} the deadline at tight cells, so it misses
$59$--$100\%$ structurally, while the aware refit holds every primary
cell under budget; on the Nano the blind policy misses $34$--$100\%$
where the aware refit holds ${\le}0.1\%$. The energy direction is
duty-dependent and the blind policy's apparent savings are illusory ---
it is infeasible: for periodic vision the aware policy's higher feasible
clock costs $+1$--$24$\,mJ/inf ($1$--$3\%$), while for saturated decode
race-to-completion~\cite{criticalpowerslope2002} makes the higher clock
$10$--$20$\,mJ/token \emph{cheaper} --- the aware model's role there is to mark which clocks
are feasible at all.

The missing state is a property of the family, not one unit. Repeating
the upper-range sweep on the second SKU, the $2133{\to}3199$\,MHz
penalty matches in magnitude and sign (NX: MobileNetV2 $+15.8\%$, GEMV
proxy $+44.5\%$, SLM $+18.7\%$; Nano: $+11.7\%$/$+44.8\%$/$+15.3\%$),
the workload ranking transfers intact, and the low-clock collapse
reaches $5$--$15\times$ on both boards --- on the matched, verified EMC
axis the two SKUs share, with surrounding platform differences still
able to shape the magnitude and manifestation of a matched step
(Supp.~\S S3).

Why a table rather than a term (Table~\ref{tab:solo}, bottom): a
parametric extension ${+}m/f_{\mathrm{emc}}$, fitted at $\{3199,
665.6\}$\,MHz and interpolated to the held-out 2133\,MHz point, is
\emph{worse than the unrepaired model} for three of four workloads, and a
monotone power-law sum in the style of prior memory--compute scaling
work~\cite{jointmemfreq2025} predicts a held-out difference with the
wrong \emph{sign} ($+2.67$\,ms predicted, $-1.70$\,ms measured). We
tabulate per lockable EMC point rather than fit a frequency term also
because the EMC response need not even be monotone: a reproducible scoped
counterexample (Nano, top GPU clock, ONNX Runtime cuBLAS: $-9.1\%$,
lower-energy per inference as well) is analyzed in Supp.~\S S6. The
two-cell refit bounds median held-out error to $2.9$--$4.6\%$ for three
of four workloads --- bounded, though not always to two cells: the
bandwidth-bound decode proxy needs the full per-point sweep, and the
repair is robust to which two cells are profiled ($26/28$ pairs hold a
${\le}10\%$ worst case on MobileNetV2; Supp.~\S S3). The sensitivity itself is no fp32-dispatch artifact
(Table~\ref{tab:trtort}): rebuilding the ONNX workloads into fused
TensorRT~\cite{tensorrt} fp16 engines and re-running the locked EMC
sweep, the penalties roughly halve but persist at $+8$--$21\%$ --- an
estimator calibrated on the wrong runtime would mis-scale the EMC term
even where it correctly includes it, so the per-point table must be
built under the deployed runtime.

\begin{table}[t]
  \centering
  \caption{$2133\to3199$\,MHz median penalty (\%; positive = lower clock
  slower) under TensorRT fp16 vs.\ ONNX Runtime CUDA fp32 on the Nano, CPU
  and GPU pinned. The EMC sensitivity persists under both runtimes but is
  roughly halved by the fused engine; the L2-resident GEMM's sign flips ---
  the inversion is present only under ONNX Runtime.}
  \label{tab:trtort}
  \resizebox{\columnwidth}{!}{%
  \begin{tabular}{lrrl}
    \toprule
    workload & TensorRT fp16 & ORT fp32 & note \\
    \midrule
    MobileNetV2      & $+9.9$  & $+11.7$ & monotone, both \\
    ViT-Small        & $+7.8$  & $+18.0$ & monotone, both \\
    GEMV proxy       & $+20.8$ & $+44.8$ & monotone, both \\
    GEMM L2-resident & $+8.1$  & $-9.1$  & inversion: ORT only \\
    \bottomrule
  \end{tabular}}
\end{table}
The repair is deliberately simple --- a two-cell refit at the deployment
memory clock, not additional controller machinery --- and that is the point: what
changes deployed feasibility is the missing state, not controller complexity.
The missing variable is bounded and directly measurable; omitting it is what sends the
governor to an infeasible operating point.

\begin{figure*}[t]\centering\includegraphics[width=\textwidth]{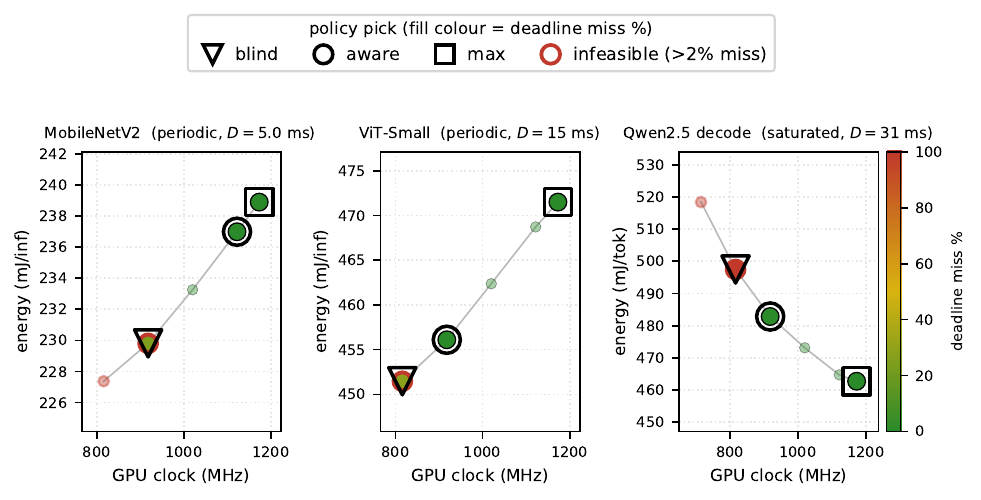}
\caption{\textbf{A deployed governor's feasible-energy frontier across three workload classes} (NX, calibrated $1$\,ms-sampled module-rail energy; point colour $=$ measured deadline-miss \%, warmer $=$ higher miss; points above the $2\%$ deployment miss budget are outlined in red and \emph{infeasible}). At a tight deadline the EMC-blind GPU-only fit selects an \emph{infeasible} clock in every case; the EMC-aware fit selects a feasible one. The energy-optimal feasible point is \emph{duty-dependent}: for periodic vision (idle gaps, pinned clock) energy \emph{rises} with clock, so the lowest feasible clock is near-optimal and the maximum is wasteful; for saturated LLM decode \emph{race-to-completion} makes energy \emph{fall} with clock, so the frontier instead rewards running fast (\S\ref{sec:solo}). Bold dots are the three policy picks; faint dots are the other measured operating points. Each policy selects by its prediction-plus-margin rule ($T{+}M{\le}D$), so the aware pick need not be the lowest-energy feasible dot.}\label{fig:frontier}\end{figure*}

The energy axis is the payoff, and its shape is duty-dependent
(Fig.~\ref{fig:frontier}). For periodic vision the deployment pins the
selected clock, so the aware policy's higher feasible clock draws
slightly more energy ($+1$--$24$\,mJ/inf, $1$--$3\%$) --- the blind
policy's apparent savings are illusory because it is infeasible, and a
deadline-blind max-clock fallback wastes up to ${\sim}9$\,mJ/inf over
the aware pick. For saturated decode, race-to-completion inverts the
ordering: per-token energy falls monotonically with clock, so the aware
policy is feasible \emph{and} $10$--$20$\,mJ/token cheaper than the
infeasible blind choice, and the energy-minimal feasible point is the
\emph{maximum} clock. The EMC-aware model thus converts
deadline-missing under-provisioning or wasteful over-provisioning into
energy-efficient feasible operation, with the optimal point selected by
duty.

\section{Two Further State Layers: Horizon and Tenancy}\label{sec:layers}

\begin{figure*}[t]\centering
\includegraphics[width=\textwidth]{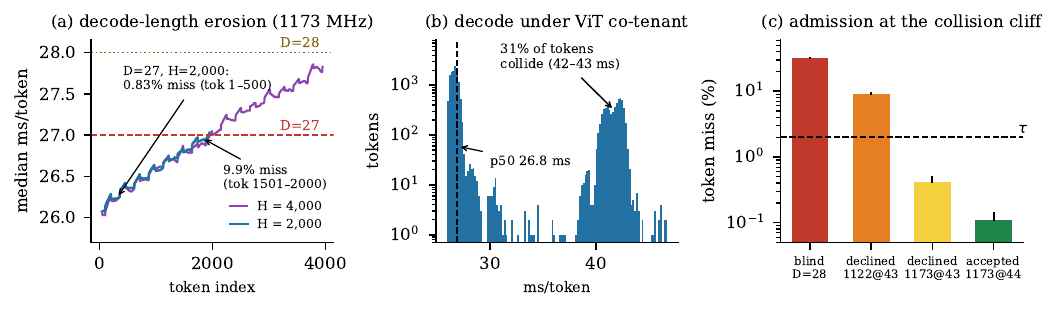}
\caption{\textbf{Horizon and tenancy as contract state} (Qwen2.5-1.5B,
NX, EMC 2133\,MHz). \textbf{(a)}~Decode-length erosion: per-token median
latency drifts upward with KV-cache growth ($+2.5$--$2.7\%$ by token
$2{,}000$), multiplying the late-window miss rate at $D{=}27$\,ms/tok
from $0.83\%$ (tokens 1--500) to $9.9\%$ (tokens 1501--2000); at
$H{=}4{,}000$ the drift ($+6.0\%$ measured, $+5.8\%$ linear
extrapolation) crosses the deadline mid-response.
\textbf{(b)}~Co-tenancy turns decode bimodal: against a 10\,Hz ViT
co-runner the median token moves $26.12{\to}26.82$\,ms while $31\%$ of
tokens collide onto a $42$--$43$\,ms plateau --- a tail no median-fit
model expresses. \textbf{(c)}~Admission outcomes at the collision cliff:
tenancy-blind admission at $D{=}28$\,ms misses $32.3\%$; at $D{=}43$\,ms
the declined 1122\,MHz cell measures $9.0\%$ (decline correct) and the
first accepted contract ($D{=}44$\,ms, 1173\,MHz) measures $0.11\%$
over nine runs.}
\label{fig:layers}
\end{figure*}

\textbf{Decode horizon.} Deploying Qwen2.5-1.5B token decode (greedy,
batch~1, all layers on GPU) at per-token deadlines, the aware policy's
$300$-token cells hold ${\le}0.9\%$ --- but $2{,}000$-token runs expose
\emph{decode-length erosion of feasibility}
(Fig.~\ref{fig:layers}a). Per-token median latency drifts upward as the
KV cache grows ($+2.7\%$ by token $2{,}000$ at the tightest cell), and on
a knife-edge distribution that small drift multiplies the miss rate:
tokens $1$--$500$ miss $0.83\%$ while tokens $1{,}501$--$2{,}000$ miss
$9.9\%$ at $D{=}27$\,ms/tok (eight runs), crossing the budget --- and at that deadline
\emph{no} clock is feasible for long windows, since 1173\,MHz is already
the maximum. Feasibility for saturated decode is therefore a property of
the $(D, \text{decode-horizon})$ pair, not of the clock alone: a governor
must provision for the horizon or re-actuate as context grows. The
exposure is systematic across model size, quantization, and context
(Supp.~\S S2); Table~\ref{tab:qwengov} gives the deployed cells, and the
admission consequence is what matters here.

\begin{table}[ht]\centering\footnotesize
\caption{Qwen2.5-1.5B token-decode deployed governor on the NX (EMC 2133\,MHz; per-token TPOT SLO). miss\%: three $300$-token repeats, and --- because KV-cache
growth drifts per-token latency upward --- five $2{,}000$-token repeats (eight at the $D{=}27$ cell)
[run-level 95\% upper bound]. Calibrated $1$\,ms-sampled energy; saturated
decode, so $\Delta E<0$ means aware \emph{lower} via race-to-completion.}
\label{tab:qwengov}
\begin{tabular}{@{}lrrrrrr@{}}\toprule
& \multicolumn{2}{c}{blind (@3199)} & \multicolumn{3}{c}{aware (@2133)} & $\Delta E$\\
\cmidrule(lr){2-3}\cmidrule(lr){4-6}
$D$\,(ms/tok) & $f^\star$ & miss\% & $f^\star$ & @300 & @2k [UB] & mJ/tok\\\midrule
27 & 1020 & $100$ & 1173 & $0.9$ & $3.2$ [$3.6$]$^{\dagger}$ & $-10$\\
29 & 918 & $100$ & 1020 & $0.7$ & $0.8$ [$0.8$] & $-10$\\
31 & 816 & $100$ & 918 & $0.8$ & $0.7$ [$0.8$] & $-14$\\
\bottomrule\end{tabular}
\end{table}

\textbf{Co-tenancy.} Running MobileNetV2 against a rate-limited ViT-Small
co-runner (10\,Hz, a second process on the same GPU), the victim's
\emph{median} barely moves --- it even improves ${\sim}7\%$ at the top
clock, the co-runner keeping the GPU out of its idle states --- but
$10$--$14\%$ of cycles now queue behind a co-runner inference, opening a
$4$--$7$\,ms tail: at $\tau{=}2\%$, $D{=}7$\,ms is infeasible at
\emph{every} clock, and $D{=}9$\,ms only at ${\ge}1020$\,MHz
($0.0$--$0.2\%$ vs.\ $10.1\%$ at 918\,MHz). No median-fit policy ---
blind, EMC-aware, or even one re-profiled \emph{under} the co-runner ---
can express this: the medians it fits still admit clocks whose tails miss
$10\%$ of deadlines. Tenancy does not invalidate the per-EMC-point table
(the median model stays accurate, and the profile-pinned EMC holds
exactly under two GPU clients --- $165/165$ half-second samples); it adds
a tail dimension to the feasibility test: the margin must be conditioned
on occupancy rather than on the solo knife-edge distribution alone. For
decode the same co-runner is more violent still: the token distribution
turns \emph{bimodal}, the median moving $26.12{\to}26.82$\,ms while
$31\%$ of tokens collide onto a $42$--$43$\,ms plateau
(Fig.~\ref{fig:layers}b) --- motivating the composed admission rule that
follows.

\section{The Guarded Contract-Admission Policy}\label{sec:policy}

The layers compose into a single deployed pick rule, each restored by a
bounded probe: the two-cell refit gives the per-EMC-point model
$\hat{T}(f)$; the measured decode drift gives a per-token coefficient
$\gamma$; and a $200$-cycle probe at the top clock \emph{under the
co-runner} gives an occupancy margin $M$. A contract $(D, H)$ is feasible
at clock $f$ when
$\hat{T}(f)\,(1{+}\gamma H)\,r + M\,f_{\mathrm{probe}}/f \le D$; the
policy selects the lowest feasible clock and otherwise declares the
contract infeasible. The $f_{\mathrm{probe}}/f$ scaling is the
first-order statement that the co-runner's kernels --- and with them the
queueing quantum a victim release can land behind --- dilate as the
shared GPU slows. It is an analytic model, not a fitted exponent: its
single coefficient is fixed by the top-clock probe, no evaluation miss
rate is used to tune it, and the probe and all evaluation runs are
disjoint. The composed controller is accordingly a
\emph{contract-admission} policy, not a cold-start one: it spends a
bounded probe (200 cycles, ${\sim}6$\,s at 33\,Hz) characterizing the
active deployment condition, then selects the operating point before the
admitted contract's first production release. We keep the vocabulary
strict: a contract is \emph{predicted-admissible} when the rule selects a
clock for it, and \emph{decisively feasible} when the measured run-level
$95\%$ upper bound sits below the budget.
Here $r$ is the same fixed solo-condition safety multiplier
($1.035$, \S\ref{sec:solo}), $\gamma$ denotes fractional latency
growth per decoded token, and $H$ is the requested post-warm-up decode
horizon in tokens. The probe is measured once per stable deployment
condition; the independent runs of \S\ref{sec:account} validate the
selected contract's execution under that condition, not repeated
probing.

Deployed live on the NX (MobileNetV2 at 33\,Hz against the 10\,Hz
ViT-Small co-runner, EMC locked at 2133\,MHz), the probe measures
$M{=}3.52$\,ms --- within $0.01$\,ms of the value the independent
co-tenancy traces above imply. Table~\ref{tab:ladder} scores each policy at its own pick over $3{\times}1{,}000$ cycles per cell. The
solo-margin policies choose 510\,MHz at both deadlines --- blind and
EMC-aware alike, because the solo medians admit it --- and miss $24.3\%$
($D{=}9$\,ms) and $20.4\%$ ($D{=}12$\,ms) of deadlines: under
co-tenancy, repairing the point model alone repairs nothing. Applying
the probe margin \emph{flat} --- the variant we deployed first ---
repairs $D{=}9$\,ms (it selects 1020\,MHz; $2.5\%$ measured, at the
budget edge) but still admits 510\,MHz at $D{=}12$\,ms and misses
$20.4\%$: a margin probed at the top clock understates the tail
everywhere below it, because the occupancy quantum is itself a function
of the candidate clock, not a constant of the co-runner. The
clock-scaled rule selects 1122\,MHz at $D{=}9$\,ms --- decisively
feasible ($0.43\%$; run-level $95\%$ upper bound $0.70\%$) --- and
816\,MHz at $D{=}12$\,ms, a knife-edge cell we report as \emph{at the
budget boundary}, not met: the mean, $1.90\%$, sits just inside, but the
run spread ($0.8$--$2.7\%$) puts the run-level upper bound at $2.53\%$.
One admitted step of headroom, 918\,MHz, is decisively inside
($0.22\%$; run-level upper bound $0.40\%$ over ten independent runs). By measured mean the rule's feasibility
classification agrees in all $14$ deployed (clock, deadline) cells, with
that single boundary cell statistically unresolved at run level.

The
probe itself supplies a principled guard. Under an i.i.d.\ reference calculation~\cite{wilks1941}, only the sample
\emph{maximum} provides at least $95\%$ coverage of the probe
distribution's $98$th percentile from $200$ samples
($0.98^{200}{\approx}0.018$; the second-largest order statistic
reaches only $91\%$); because collisions are weakly and
quasi-periodically correlated within a probe (the decode probe's
autocorrelation oscillates in sign, reaching $+0.63$ at lag $3$), we use
it solely as a \emph{conservative empirical guard}, not as a formally
calibrated confidence guarantee. The guarded rule
therefore admits with $M_{\mathrm{g}} = \max - p50$ of the probe
($4.36$ vs.\ $3.52$\,ms) --- deterministic, probe data only, no
evaluation feedback. The maximum rule avoids relying on the unstable
cycle-level bootstrap --- whose coverage trap under clustered exceedances
the margin-calibration analysis of \S\ref{app:gpd} documents, and whose $p98$
resample distribution on a $200$-sample probe is too discrete to yield a
stable $95\%$ point --- while remaining, under correlated collisions, an
empirically conservative guard rather than a calibrated coverage
bound. The guard promotes the $D{=}12$\,ms pick to
918\,MHz, decisively feasible, and leaves the joint-stress acceptance
below unchanged, at the cost of declining $D{=}9$\,ms --- whose
point-estimate pick is in fact measured-feasible. The guarded rule
trades utilization for decisiveness: every contract it accepts in this
evaluation is decisively feasible.

The same rule prices the decode horizon. Inverting the feasibility test
gives a quoted horizon cap $H^{*}(f) = (D/(r\,\hat{T}(f)) - 1)/\gamma$:
at $D{=}28$\,ms per token, $H^{*}{\approx}2{,}140$ tokens at
1173\,MHz. The quote separates two contracts the horizon-blind rule
accepts identically: a $2{,}000$-token response (measured $0.52\%$ miss
over eight runs --- feasible) and a $4{,}000$-token one, which the horizon-blind rule
schedules at its cheapest admitted clock, 1122\,MHz, where the drift
crosses the deadline mid-response --- the last $500$ tokens miss
$100\%$ ($21.3\%$ overall). At $H{=}2{,}000$ the ${\sim}2.5\%$ drift
this clock accumulates sits inside the static $3.5\%$ margin, so
horizon-blind acceptance survives by accident; by $H{=}4{,}000$ the
measured drift reaches $+6.0\%$ --- matching linear extrapolation of the
$2{,}000$-token coefficient to within $0.2$ percentage points --- and
the margin is spent twice over. The horizon layer converts that
accident into an explicit, quoted bound. The rule \emph{conservatively
declines} the $4{,}000$-token contract under its linear admission model
rather than proving it infeasible: at the maximum clock the declined
contract measures $1.87\%$ over its last $500$ tokens --- inside the
budget with no headroom left --- while the horizon-blind policy admits it
at the cheaper 1122\,MHz and fails catastrophically late in decode.

The two contracts above exercise the layers separately --- the co-tenant
deployment carries no horizon term, the decode contracts no co-tenant ---
so we also deploy the joint-stress cell: $2{,}000$-token Qwen decode
against the same 10\,Hz ViT co-runner, all three layers active in one
workload. Under the co-tenant the decode distribution turns bimodal: the
median token barely moves ($26.12{\to}26.82$\,ms, $+2.7\%$), but $31\%$
of tokens collide with a co-runner inference and land on a plateau near
$42$--$43$\,ms. A tenancy-blind horizon-aware rule, whose solo model
accepts $D{=}28$\,ms per token, therefore misses $32.3\%$ of tokens ($31.9$--$32.6\%$ over nine runs) ---
sixteen times its budget. The composed rule re-probes under the
co-tenant ($200$ tokens, ${\sim}5$\,s): the probe returns a
barely-shifted level ($+0.2\%$) and $M{=}15.8$\,ms of occupancy tail,
pushing the first accepted contract to $D{=}44$\,ms at 1173\,MHz ---
measured $0.11\%$ (run-level $95\%$ upper bound $0.14\%$ over nine
independent $2{,}000$-token deployments), decisively feasible.
Its refusals bracket the collision cliff from the safe side: at
$D{=}43$\,ms it declines 1122\,MHz, which measures $9.0\%$, and declines
the top clock, which measures $0.41\%$ --- conservative by one cell, at
a boundary where one deadline millisecond separates $9\%$ from $0.2\%$.
The solo drift coefficient transfers: measured drift under the co-tenant
is $+2.3$--$2.5\%$ per $2{,}000$ tokens against $+2.5$--$2.6\%$ solo, so
the horizon term needs no re-probe.

A reactive backstop complements admission for drift the probe cannot
foresee, and we verify it live. With an EMC actuation lag
$L{\approx}8$\,ms, same-cycle rescue is impossible at every deadline
deployed here --- the inference that triggered the switch completes
under the old clock --- but next-release actuation is feasible whenever
post-completion slack exceeds the lag. Running MobileNetV2 at an
infeasible cell (GPU 918\,MHz, $D{=}5$\,ms, sustained miss
$38$--$55\%$), a controller that raises the EMC to 3199\,MHz on the
first miss reads back the new clock by the next release and recovers
within $1$--$3$ cycles at a cost of at most one additional miss (three
repeats): online memory-clock adaptation is practicable one period
late, never same-cycle.
Two caveats keep the composition honest. The repeat spread at the
$D{=}9$\,ms knife-edge ($0.1$--$4.2\%$ across the three 1020\,MHz runs)
shows the occupancy tail itself moves between runs --- drift of exactly
the kind the first-miss upshift backstop demonstrated above exists to
catch; and the drift and margin models are deliberately first-order,
evaluated for two victim workloads against one co-runner type. The
occupancy probe is likewise valid for the observed co-tenancy condition:
a material change in co-runner identity or rate during the contract
requires re-probing or a conservative reactive backstop, which we do not
evaluate here.

\section{End-to-End Accounting}\label{sec:account}

Table~\ref{tab:admission} closes the efficiency question
a guard invites --- how many genuinely feasible contracts it turns away.
Across the seven evaluated contracts, the guarded policy makes no false
admission under the run-level criterion, at the cost of conservative
refusals near the feasibility boundary: of its four refusals, one is
measured-correct (every candidate clock misses ${\ge}32\%$), one
declines a contract that is itself unresolved at window level (last-500
upper bound $2.27\%$), and the two fully-resolved conservative refusals
sit within one deadline millisecond --- one GPU bin --- of the guarded
acceptance threshold. The guard is not the trivial always-refuse
policy: its conservatism is concentrated in the boundary band.

\begin{table}[t]
  \centering
  \caption{Admission outcomes of the guarded composed policy over the
  seven evaluated contracts. $n$ = independent deployment runs
  ($1{,}000$ cycles each for vision; $2{,}000$- or $4{,}000$-token
  decodes). For accepted contracts the measured miss is at the accepted
  pick; for declined contracts, at the strongest declined candidate.
  Parentheses give the run-level $95\%$ upper bound. This table reports the
  reference-probe accounting under the \emph{maximum guard}; the
  dispersion-banded policy is stress-tested in
  Table~\ref{tab:fresh}, and at the reference probe the band leaves
  the vision pick unchanged. No false admissions occur under this
  reference-probe accounting; refusals concentrate at the feasibility
  boundary.
  ($^\dagger$window-level unresolved: last $500$ tokens $1.87\%$,
  upper bound $2.27\%$.)}
  \label{tab:admission}
  \footnotesize
  \setlength{\tabcolsep}{3.5pt}
  \resizebox{\columnwidth}{!}{%
  \begin{tabular}{llrll}
    \toprule
    contract & outcome & $n$ & measured\,\% (UB) & verdict \\
    \midrule
    vision co, $D{=}9$\,ms   & declined & 3 & 0.43 (0.70) @1122 & conservative \\
    vision co, $D{=}12$\,ms  & 918\,MHz & 10 & 0.22 (0.40) & true admission \\
    decode, $D{=}28$, $H{=}2$k & 1173\,MHz & 8 & 0.52 (0.55) & true admission \\
    decode, $D{=}28$, $H{=}4$k & declined & 3 & 0.85 (0.92) @1173$^\dagger$ & conservative \\
    decode co, $D{=}28$ & declined & 9 & 32.3 @1173 & correct refusal \\
    decode co, $D{=}43$ & declined & 9 & 0.41 (0.49) @1173 & conservative \\
    decode co, $D{=}44$ & 1173\,MHz & 9 & 0.11 (0.14) & true admission \\
    \bottomrule
  \end{tabular}}
\end{table}

\begin{table}[t]
  \centering
  \caption{Ablation ladder under the ViT co-runner (NX, MobileNetV2 at
  33\,Hz, EMC 2133\,MHz): each policy's live pick and measured miss at
  that pick ($3{\times}1{,}000$ cycles; run range in parentheses;
  run-level $95\%$ bootstrap upper bound). Solo-margin policies pick the
  same infeasible clock whether EMC-aware or not; the flat probe margin
  repairs only $D{=}9$\,ms; the clock-scaled margin meets $D{=}9$\,ms
  decisively and accepts a knife-edge cell at $D{=}12$\,ms whose upper
  bound straddles the budget; the guard promotes it to a decisively
  feasible pick. Tenancy-blind decode admission (co-tenant,
  $D{=}28$\,ms/tok) misses $32.3\%$.}
  \label{tab:ladder}
  \footnotesize
  \setlength{\tabcolsep}{4pt}
  \resizebox{\columnwidth}{!}{%
  \begin{tabular}{llrrr}
    \toprule
    policy (margin) & $D$ & pick & miss\,\% (runs) & UB\,\%\\
    \midrule
    blind / EMC-aware (solo) & 9  & 510  & 24.3 (24.2--24.4) & 24.4\\
    ~~+ tenancy, flat $M$    & 9  & 1020 & 2.5 (0.1--4.2)    & 3.8\\
    ~~+ tenancy, $M f_{\mathrm{probe}}/f$ & 9 & 1122 & 0.43 (0.0--0.8) & 0.70\\
    \addlinespace
    blind / EMC-aware (solo) & 12 & 510  & 20.4 (18.5--21.4) & 21.3\\
    ~~+ tenancy, $M f_{\mathrm{probe}}/f$ & 12 & 816 & 1.90 (0.8--2.7) & 2.53\\
    ~~+ tenancy, maximum guard ($M_g$) & 12 & 918 & 0.22 (0.0--1.1) & 0.40\\
    \addlinespace
    tenancy-blind (decode, co) & 28/tok & 1173 & 32.3 (31.9--32.6) & 32.4\\
    \bottomrule
  \end{tabular}}
\end{table}

The point
of the deployment is correspondingly narrow and strong: with all three
state layers in its input --- each from a bounded probe --- every
guarded-policy acceptance under the reference-probe accounting is
decisively feasible at the stated run-level criterion, while the point-estimate variant separately accepts
one knife-edge vision cell that remains statistically unresolved at the
budget boundary; refusals
are either measured-correct or concentrated at the feasibility
boundary (Table~\ref{tab:admission}); and every policy missing a layer violates its budget
at a deployed contract: for every omitted state layer, at least one
measured contract shows the corresponding ablation violating the budget
by $20$--$32\%$.

The accounting above evaluates contracts admitted from one reference
probe per condition. Sixteen fresh-probe trials then
\emph{stress-test} the admission pipeline itself --- each trial
launches a fresh co-runner, takes a fresh probe, makes a fresh guarded
decision, and runs a disjoint production contract
(Table~\ref{tab:fresh}; Supp.~\S S2). The decode arm is
pipeline-robust: fresh probes returned larger margins ($M_{\mathrm{g}}$
$15.9$--$17.5$\,ms), the rule quoted $D{=}45$--$46$\,ms, and all eight
contracts measured ${\le}0.15\%$. (These trials sweep the requested
deadline to each probe realization's first predicted-admissible
contract --- characterizing the variability of the admission boundary
--- while the fixed $D{=}44$\,ms contract of
Table~\ref{tab:admission} is the reference-probe result held over nine
independent runs.) The vision arm exposes a \emph{probe-variance
failure mode}: six trials admitted 918\,MHz with ${\ge}415\,\mu$s of
feasibility slack and measured $0$--$1.4\%$, but two trials whose probe
realization returned a smaller maximum admitted the knife-edge
816\,MHz clock with only $138$--$231\,\mu$s of slack --- inside the
probe margin's own cross-trial dispersion (s.d.\ $199\,\mu$s over nine
probes) --- and measured $3.0$--$3.1\%$: false admissions waiting on an
unlucky probe.

That criterion is directly actionable, and we close the loop on it.
To keep the variants distinct: the \emph{point policy} admits on the
probe's $p98$ margin $M$; the \emph{maximum guard} on
$M_{\mathrm{g}} = \max - p50$; and the \emph{final banded policy}
adds a slack band $h$ to the maximum guard. Writing the feasibility
slack as
$s(f) = D - [\hat{T}(f)(1{+}\gamma H)\,r + M_{\mathrm{g}}\,
f_{\mathrm{probe}}/f]$, the \emph{dispersion-banded} rule admits the
lowest $f$ with $s(f) \ge h$, where $h = 2\sigma_{\mathrm{probe}} =
400\,\mu$s is a fixed (not clock-scaled) band equal to twice the probe
margin's cross-trial standard deviation --- calibrated, we state
plainly, \emph{after} observing the two failures, from the round-1
probe population alone, and fixed before a held-out validation round
(band-width sensitivity: any $h$ between $1.5\sigma$ and $2\sigma$
selects identical held-out picks; Supp.~\S S2). The band is calibrated
and evaluated on the vision arm, where the failure occurred; at the
reference probe it leaves the vision pick unchanged (918\,MHz, slack
$605\,\mu$s ${\ge}h$), and the decode arm's quotes are unbanded ---
its fresh boundary sweeps were uniformly more conservative than the
reference quote and all passed (Table~\ref{tab:fresh}).
In ten further end-to-end trials the band fired on four probes that
would otherwise have admitted 816\,MHz, no boundary clock was
admitted, and every trial selected 918\,MHz. The held-out contracts
then expose the third and final uncertainty layer: per-launch outcomes
at the knife-edge spread $0.0$--$2.9\%$ (two of ten $1{,}000$-cycle
windows above $\tau$), \emph{uncorrelated} with the trial's own probe
margin (correlation $+0.08$) --- co-runner launch and phase realization
move the knife-edge tail in ways the tested one-shot probe did not
predict. We therefore separate three
criteria the results now require: \emph{reference-probe decisive
feasibility} (run-level upper bound below $\tau$ across repeated
deployments under one probe condition; Table~\ref{tab:admission}),
\emph{fresh-launch aggregate feasibility} (mean and launch-level upper
bound below $\tau$ across independent launches), and \emph{per-launch
compliance} (every launch window below $\tau$). The banded pick
achieves the first two at the knife-edge --- across the ten held-out
banded launches, $1.19\%$ with a launch-level upper bound of $1.70\%$,
below the budget --- but not the third. (Pooling all sixteen same-clock
918\,MHz launches, including the six selected by the unbanded rule,
gives a descriptive $0.96\%$ [$1.34\%$]; we do not use this
selection-conditioned pool to validate the banded policy.) The third,
too, has a measured price: repeating the fresh-launch protocol one bin
above the banded pick (1020\,MHz, eight independent co-runner
launches), \emph{every} launch measures $0.00\%$ at $D{=}12$\,ms ---
observed per-launch compliance across all eight launches, purchased
with one step of headroom. The admission-uncertainty stack is
therefore three measured layers with three measured prices: omitted
deployment state ($20$--$32\%$ misses; repaired by bounded probes),
probe-induced low-clock admission ($3.0$--$3.1\%$ at the knife-edge;
eliminated by the dispersion band in the held-out trials), and
condition realization (per-launch spread $0$--$2.9\%$ around a
compliant aggregate, which the tested one-shot $200$-cycle probe did
not predict --- correlation $+0.08$ over ten trials --- and which one
additional bin empirically absorbed across eight fresh launches). Based
on the solo-calibrated per-clock energies of \S\ref{sec:solo}, the
$918{\to}1020$\,MHz step is estimated to add roughly $2$\,mJ per
inference (${\sim}1\%$); co-tenant energy was not separately
instrumented. The decode arm,
whose accepted contracts sit well inside their boundaries, shows none
of this.

\begin{table}[t]
  \centering
  \caption{End-to-end fresh-probe stress tests: each trial = fresh
  co-runner launch $\to$ fresh probe $\to$ fresh admission decision
  $\to$ disjoint production contract. The dispersion band ($h =
  2\sigma_{\mathrm{probe}} = 400\,\mu$s) is calibrated on the round-1
  probes after observing the two failures and validated held-out. Zero
  misses were observed in all eight $1{,}000$-cycle launches at
  1020\,MHz; because cycles are serially correlated, we do not
  interpret pooled zero events as a calibrated binomial bound.}
  \label{tab:fresh}
  \footnotesize
  \setlength{\tabcolsep}{3.5pt}
  \resizebox{\columnwidth}{!}{%
  \begin{tabular}{lllllll}
    \toprule
    arm & trials & pick & contract & mean [UB] & range & outcome \\
    \midrule
    vision, single-probe & 6/8 & 918 & $D{=}12$ & --- & 0--1.4 & safe \\
    vision, single-probe & 2/8 & 816 & $D{=}12$ & --- & 3.0--3.1 & \textbf{false admission} \\
    vision, banded (held-out) & 10/10 & 918 & $D{=}12$ & 1.19 [1.70] & 0--2.9 & aggregate-feasible \\
    vision, one-bin headroom & 8/8 & 1020 & $D{=}12$ & 0.00 observed & 0--0 & observed per-launch compliant \\
    decode co, boundary sweep & 8/8 & 1173 & first-adm.\ $D{=}45$--$46$ & 0.06 [0.09] & 0--0.15 & robust \\
    \bottomrule
  \end{tabular}}
\end{table}

Deployed estimators are not fully static, and the comparison credits
that: we \emph{deploy} a FLAME-style~\cite{flame2026} EWMA drift calibrator
($\alpha{=}0.1$, no hysteresis, re-picking every cycle) rather than argue
about it. Its converged effect is a constant offset that re-centres
median held-out bias below $3\%$ --- but a level offset structurally
cannot repair the memory-clock-dependent \emph{slope}, so the worst-case
error stays $20$--$29\%$ for the memory-bound workloads (decode proxy
$28.6\%$, ViT $11.4\%$). Live at the tightest deadline it never settles,
crossing the knife-edge between the top two clocks $13$--$15$ times per
$2{,}000$-cycle run --- an oscillation its stationary replay does not
predict, and one insensitive to the gain ($5$--$73$-cycle escapes across
$\alpha \in \{0.02, 0.1, 0.3\}$). The honest concession: for level-shift
failures it escapes fast, within $1$--$8$ cycles at $0$--$6$ transient
misses. But for LLM decode the level itself is non-stationary --- KV
growth drifts the target the mean-tracker chases --- and the EMC-aware
policy pays none of these: right operating point from the first release,
zero transient, zero hunting, no gain to tune.

\section{Limitations}\label{sec:limits}

The study spans two boards of one family: both are integrated,
BPMP-class NVIDIA SoCs sharing one LPDDR5 controller and the same four
lockable EMC points, so we scope our claims to that class and release
the harness so other parts can be profiled directly; surrounding
hardware and software differences can still affect the magnitude and
qualitative manifestation of the response to a matched EMC step. The
composed policy is deliberately first-order --- two victim workloads, one
co-runner type, linear drift and margin models --- not a production
governor; its role is to demonstrate that the identified
deployment-state categories \emph{suffice} for proactive admission. The
occupancy probe is valid for the observed co-tenancy condition: a
material change in co-runner identity or rate during the contract
requires re-probing or a conservative reactive backstop, which we do not
evaluate here. The dispersion band is calibrated from one condition's probe
population after observing the round-1 failures and validated on
held-out trials of the same condition; its transfer to other workloads
and co-runners is not evaluated. At the knife-edge contract, per-launch
outcomes vary $0$--$2.9\%$ around a compliant aggregate; the tested
one-shot $200$-cycle probe did not predict this variation ($n{=}10$,
correlation $+0.08$) --- longer, repeated, or phase-aware probes might
--- and per-launch compliance was obtained with one bin of headroom
under eight launches of the one evaluated co-runner
(\S\ref{sec:account}). The one-bin headroom experiment is a measured
design point, not a learned or universally calibrated rule for
selecting a per-launch guarantee tier. The probe margin is likewise conditioned at the probed
EMC point, and $f_{\mathrm{probe}}/f$ scales only the GPU-clock dilation
of the co-runner quantum --- growth in collision \emph{probability} at
low clocks is second-order and absorbed only by the guard's conservatism
near the probe clock. Admission enforces an aggregate miss budget, not a
weakly-hard pattern bound: misses cluster (continuation probability up to
$0.74$, $360$--$740\times$ independence), characterized in
Supp.~\S S5. Sustained operation is certified from $1{,}000$-cycle
deployments plus a one-hour saturated soak; ambient temperature was
recorded, not controlled.

\section{Conclusion}\label{sec:conclusion}

The memory clock is a deployment state that edge-inference deadline
governors must see. On two integrated Orin SoCs, a CPU$\times$GPU
frequency-aware latency estimator --- valid within its profiling scope
--- is calibrated to the mean of a distribution whose location, shape,
and actuation all depend on the EMC axis it does not observe; deployed,
that omission sends the governor to operating points that miss
$25$--$28\%$ of deadlines where an EMC-aware refit holds ${\le}0.9\%$,
across six models on two SKUs. The deployment state a governor needs is
layered: EMC state repairs the \emph{point model}, decode horizon
updates that model \emph{over time}, and under the tested co-tenancy
condition occupancy can change the \emph{tail margin} required around it
--- a \emph{latency-only} reactive controller can infer these effects
only after observing responses, where a governor supplied with EMC,
horizon, and tenancy state has the information needed for proactive
selection --- and we close that loop by deploying one as a
\emph{contract-admission} policy: bounded probes restore each layer,
and, under the reference-probe maximum-guard accounting, every accepted
contract across the seven evaluated contracts is decisively
measured-feasible --- including
the joint-stress cell, $2{,}000$-token decode against an active GPU
co-tenant, where tenancy-blind admission misses $32\%$ of tokens and the
composed rule's first accepted deadline measures $0.11\%$ over nine
independent runs. The
point-estimate variant separately accepts one knife-edge vision cell
that remains statistically unresolved at the budget boundary; the
guard's refusals include one measured-correct rejection and conservative
boundary refusals. End-to-end stress tests plus held-out banded trials decompose the
remaining admission uncertainty into measured layers with measured
prices: probe-induced low-clock admission, eliminated by a
dispersion-banded guard in held-out trials, and co-tenant launch
realization --- which the tested one-shot probe did not predict, and
which one additional bin of admission headroom empirically absorbed
($0.00\%$ across eight fresh launches). At knife-edge contracts the
policy's guarantee is a conservatism ladder --- aggregate compliance at
the banded pick, observed per-launch compliance one bin above ---
rather than a single safety claim. The contribution is not that reactive control is
incapable --- it is that these states belong in the governor's
\emph{input}. We release the harness and all raw data so that the EMC
axis can be folded into the next generation of edge deadline governors.
The platform-specific mechanism is Jetson's, but the systems lesson is
broader: a deadline governor cannot safely optimize over a state
abstraction that omits an independently governed shared-memory fabric ---
this is a state-abstraction failure in deployed edge-inference
governors, not a Jetson-tuning result.

\section*{Reproducibility}
The measurement harness, campaign runners, analysis scripts, and all raw
per-cycle traces are released at the artifact repository accompanying
this paper (\url{https://github.com/dankang21/jetson-latency-lab});
bulk traces are archived on Zenodo
(DOI: \href{https://doi.org/10.5281/zenodo.21236659}{10.5281/zenodo.21236659}). Appendices S1--S12 of this version contain the full measurement
study.

\clearpage
\setcounter{section}{0}
\setcounter{table}{0}
\setcounter{figure}{0}
\renewcommand{\thesection}{S\arabic{section}}
\renewcommand{\thetable}{S\arabic{table}}
\renewcommand{\thefigure}{S\arabic{figure}}
\section*{Extended Material (Appendices S1--S12)}
The following appendices provide the extended measurement study and
reproduce the claim-adjacent analyses referenced throughout the main
text, making this technical report self-contained; pointers of the form
``Supp.~\S S$n$'' in the main text resolve to these sections.

\section{Pitfalls that corrupt clock measurements}\label{sec:pitfalls}

Six failure modes produced, or would have produced, data that looked
publishable and was wrong; we report each as symptom, mechanism,
detection, and remedy.

\textbf{(P1) RT throttling injects phantom stalls.} Busy-waiting
\texttt{SCHED\_FIFO} probes intermittently report 46--51\,ms outliers
indistinguishable from genuine clock-transition stalls, because the
kernel's real-time throttling budget (\texttt{sched\_rt\_runtime\_us},
default 950\,ms per 1\,s period~\cite{linuxrtdoc}) suspends a spinning
FIFO thread once per period. Our noise-floor control windows --- identical
probes with no frequency change --- showed the same artifacts, which no
transition effect can produce; the remedy is to set
\texttt{sched\_rt\_runtime\_us} to $-1$ for the run.

\textbf{(P2) Silent BPMP rate rounding.} A cell labeled 1600\,MHz runs at
2133\,MHz while every write appears to succeed, because the BPMP rounds
off-table requests up to the next lockable rate (main text \S III). The
launch pre-flight locks and reads back every rate in the matrix and each
cell re-asserts the readback; any mismatch invalidates the cell.

\textbf{(P3) The bandwidth manager silently overrides the EMC lock.}
Symptom: \texttt{mrq\_rate\_locked} is asserted and the target written to
\texttt{rate}, every write succeeds, yet the EMC runs at a
\emph{demand-driven} rate that tracks the workload rather than the locked
value. Mechanism: on Tegra the EMC frequency is arbitrated by a BPMP
bandwidth manager that continuously re-derives a rate from observed memory
demand; \texttt{mrq\_rate\_locked} alone does not silence it, and it
overrides the lock back to its own choice. The lock takes hold only when
the bandwidth manager is halted (\texttt{bwmgr\_halt} asserted) --- which
is why \texttt{jetson\_clocks} appears to lock the EMC reliably: it
asserts \texttt{bwmgr\_halt} implicitly, a side effect easy to inherit by
accident and easy to lose when locking the EMC directly. Detection: the
trap is that the BPMP \texttt{rate} readback file reports the
\emph{requested} value and looks correct, so the override is invisible in
the obvious place to check; we caught it only because the tegrastats
\texttt{EMC\_FREQ} field disagreed with the readback under load. Remedy:
assert \texttt{bwmgr\_halt} alongside \texttt{mrq\_rate\_locked}, and
verify the lock against the tegrastats clock trace --- never against the
BPMP readback file alone (cf.\ P2, and the readback-lag finding of
\S\ref{sec:rq3}).

\textbf{(P4) stdio flushes inside timed loops.} A reproducible 4.8\,ms
``stall'' --- roughly $48\times$ our pre-specified 100\,$\mu$s relevance
criterion --- recurred at fixed cycle intervals because per-cycle logging
fills a 4\,MB stdio buffer and the flush lands inside the timed region;
its period tracked the logging volume, not the clocks. We accumulate
samples in preallocated memory and write only after the timed loop ends.
Throughout, the unit of replication is the independent deployment run ---
within-run cycles are serially correlated (\S\ref{sec:rq2}), so
run-level repeats, not within-run binomial intervals, are the honest error
bar; $\pm$ denotes the s.d.\ (or, where labeled, min--max) across such runs,
and for zero-event cells we report the observed zero misses across
the stated independent runs --- we do not treat pooled correlated
cycles as a calibrated binomial confidence bound.

\textbf{(P5) \texttt{mlockall} ordering versus CUDA.} CUDA initialization
fails with allocation errors in an otherwise correct real-time setup:
with \texttt{MCL\_FUTURE} in force, the large mappings the CUDA driver
creates at initialization must be locked and on Tegra these allocations
fail. The failure is loud, but dropping \texttt{mlockall} reintroduces
page-fault noise; instead, initialize the CUDA context first, then lock.

\textbf{(P6) Silent CPU-EP fallback.} Symptom: none --- runs complete and
write plausible latency distributions. In our configuration onnxruntime
sessions completed on the CPU execution provider, without error, when the
CUDA provider failed to initialize, caught only by a magnitude check
against known-GPU latencies after the data exists. We query the session's
resolved providers at startup and hard-fail on any mismatch.

P1 and P4 apply to any Linux real-time measurement loop, P2, P3, and P5 to
the Tegra family, and P6 to any onnxruntime deployment.

\section{Deployed-governor detail}
The directed profile-pair enumeration, per-model deployment tables, the
estimator/offset comparison, the LLM scaling axes, and the co-tenancy
policy ladder backing the summaries in main text \S IV and \S VII;
the deployed online-calibrator detail follows the tables.

\begin{table}[ht]\centering\footnotesize
\caption{Blind-policy failure across all six directed profile pairs
(MobileNetV2, NX; fit at source EMC, deploy at target EMC; deadline near the
target floor; three $1{,}000$-cycle repeats).}
\label{tab:pairs}
\begin{tabular}{@{}lrrrrr@{}}\toprule
fit$\to$deploy & $D$(ms) & blind $f^\star$ & miss\% & aware $f^\star$ & miss\%\\\midrule
3199$\to$2133 & 5.0 & 918 & $24.8$ & 1122 & $0.07$\\
2133$\to$665.6 & 8.6 & 612 & $100$ & 1173 & $0.53$\\
3199$\to$665.6 & 8.6 & 612 & $100$ & 1173 & $0.53$\\
2133$\to$3199 & 4.5 & 1173$^{o}$ & $0.0$ & 1122 & $0.00$\\
665.6$\to$3199 & 4.5 & 1173$^{o}$ & $0.0$ & 1122 & $0.00$\\
665.6$\to$2133 & 5.0 & 1173$^{o}$ & $0.1$ & 1122 & $0.07$\\
\bottomrule\end{tabular}

\smallskip\raggedright\footnotesize $^{o}$over-provisioned: feasible but the
aware refit certifies a lower-energy feasible clock.
\end{table}

\begin{table}[ht]\centering\footnotesize
\caption{ViT-Small deployed governor on the NX (EMC 2133\,MHz, MobileNetV2's
counterpart of main-text Table~II; miss\% over three $1{,}000$-cycle
repeats; $\Delta E$ = aware$-$blind calibrated run-mean module-rail energy per
inference, $n{=}2$ energy repeats per cell).}
\label{tab:vitgov}
\begin{tabular}{@{}lrrrrr@{}}\toprule
& \multicolumn{2}{c}{blind (@3199)} & \multicolumn{2}{c}{aware (@2133)} & $\Delta E$\\
\cmidrule(lr){2-3}\cmidrule(lr){4-5}
$D$\,(ms) & $f^\star$ & miss\% & $f^\star$ & miss\% & mJ/inf\\\midrule
13.6 & 918 & $100$ & 1122 & $0.1$ & $+12.6$\\
14.2 & 816 & $100$ & 1020 & $0.0$ & $+10.9$\\
15.0 & 816 & $28.0{\pm}10.7$ & 918 & $0.2$ & $+4.6$\\
\bottomrule\end{tabular}
\end{table}

\begin{table}[ht]\centering\footnotesize
\caption{The EMC exposure of LLM decode across three axes (NX, GPU
1173\,MHz, 300-token decode, three repeats; penalty $=$ per-token p50 at
2133 vs.\ 3199\,MHz).}
\label{tab:llmaxes}
\begin{tabular}{@{}llrrr@{}}\toprule
axis & config & ms/tok @2133 & penalty & mJ/tok \\\midrule
size & Qwen2.5-0.5B Q4 & $10.3$ & $+22.7\%$ & $149$ \\
     & Qwen2.5-1.5B Q4 & $26.1$ & $+18.9\%$ & $455$ \\
     & Qwen2.5-3B Q4   & $44.3$ & $+16.7\%$ & $883$ \\
quant & 1.5B Q8\_0     & $36.6$ & $+28.8\%$ & --- \\
context & 1.5B Q4, 4k prompt & $27.8$ & $+20.4\%$ & --- \\
\bottomrule\end{tabular}
\end{table}

\begin{table}[ht]\centering\footnotesize
\caption{Breadth deployments on the NX (EMC 2133\,MHz, same protocol as
main-text Tables~II and~V: blind fit @3199, aware refit
@2133, three repeats; miss\% mean over repeats; calibrated run-mean
module-rail energy, $n{=}3$ energy repeats per cell). Vision rows:
1{,}000-cycle deployments; Qwen-3B rows: 300-token
decode (five $2{,}000$-token repeats confirm the $D{\ge}47.3$ cells,
upper bounds ${\le}0.8\%$).}
\label{tab:breadthgov}
\resizebox{\columnwidth}{!}{%
\begin{tabular}{@{}llrrrrr@{}}\toprule
& & \multicolumn{2}{c}{blind (@3199)} & \multicolumn{2}{c}{aware (@2133)} & $\Delta E$\\
\cmidrule(lr){3-4}\cmidrule(lr){5-6}
workload & $D$ & $f^\star$ & miss\% & $f^\star$ & miss\% & mJ/inf\\\midrule
ResNet-50 & 13.0\,ms & 1020 & $100$ & 1173 & $0.03$ & $+17.3$\\
ResNet-50 & 14.0\,ms & 918 & $95.8{\pm}1.1$ & 1020 & $0.07$ & $+11.8$\\
ResNet-50 & 15.0\,ms & 816 & $95.5{\pm}1.2$ & 918 & $0.00$ & $+11.2$\\
YOLOv8s & 29\,ms & 918 & $100$ & 1122 & $0.00$ & $+48.7$\\
YOLOv8s & 30\,ms & 918 & $86.3{\pm}1.5$ & 1020 & $0.00$ & $+24.0$\\
YOLOv8s & 32\,ms & 816 & $59.0{\pm}9.5$ & 918 & $0.03$ & $+23.5$\\
Qwen-3B & 47.3\,ms/tok & 1020 & $33.7$ & 1122 & $0.67$ & $-11.8$ (/tok)\\
Qwen-3B & 49.3\,ms/tok & 918 & $100$ & 1020 & $0.67$ & $-19.5$ (/tok)\\
\bottomrule\end{tabular}}
\end{table}

\begin{table}[t]
  \centering
  \caption{Held-out estimator error at EMC 2133\,MHz (median/max \%): the GPU-only blind fit, a FLAME-style online drift calibrator (its converged effect is a constant offset), and the per-EMC-point table. The offset re-centres the median but cannot repair the memory-clock-dependent slope, so the worst-case error stays large for the memory-bound workloads.}
  \label{tab:offset}
  \footnotesize
  \begin{tabular}{lrrr}
    \toprule
    workload & blind & offset (drift) & EMC table \\
    \midrule
    MobileNetV2      & 5.3 / 12.8 & 0.8 / 6.1  & 2.9 / 3.4 \\
    ViT-Small        & 5.1 / 17.6 & 1.1 / 11.4 & 4.6 / 6.7 \\
    decode proxy     & 3.4 / 32.2 & 3.0 / 28.6 & 12.7 / 20.9 \\
    GEMM L2-resident & 1.5 / 2.8  & 1.6 / 2.5  & 2.9 / 4.8 \\
    \bottomrule
  \end{tabular}
\end{table}

\begin{table}[t]
  \centering
  \caption{Joint-governor deployment under the ViT co-runner (NX,
  MobileNetV2 at 33\,Hz, EMC 2133\,MHz): each policy's live pick, the
  measured miss rate at that pick with per-run range, and the run-level
  $95\%$ bootstrap upper bound ($3{\times}1{,}000$ cycles,
  $\tau{=}2\%$). Solo-margin policies pick the same infeasible clock
  whether EMC-aware or not; the flat probe margin repairs only
  $D{=}9$\,ms; the clock-scaled margin meets $D{=}9$\,ms decisively and
  accepts a knife-edge cell at $D{=}12$\,ms whose upper bound straddles
  the budget (see text).}
  \label{tab:jointgov}
  \footnotesize
  \begin{tabular}{llrrr}
    \toprule
    policy (margin) & $D$ & pick & miss\,\% (runs) & UB\,\% \\
    \midrule
    blind / EMC-aware (solo) & 9  & 510  & 24.3 (24.2--24.4) & 24.4 \\
    ~~+ tenancy, flat $M$    & 9  & 1020 & 2.5 (0.1--4.2)    & 3.8 \\
    ~~+ tenancy, $M f_{\mathrm{probe}}/f$ & 9 & 1122 & 0.43 (0.0--0.8) & 0.70 \\
    \addlinespace
    blind / EMC-aware (solo) & 12 & 510  & 20.4 (18.5--21.4) & 21.3 \\
    ~~+ tenancy, flat $M$    & 12 & 510  & 20.4 (18.5--21.4) & 21.3 \\
    ~~+ tenancy, $M f_{\mathrm{probe}}/f$ & 12 & 816 & 1.90 (0.8--2.7) & 2.53 \\
    ~~+ tenancy, guarded ($M$ UB) & 12 & 918 & 0.20 (0.0--0.4) & 0.33 \\
    \bottomrule
  \end{tabular}
\end{table}

We also \emph{deploy} the online calibrator rather than argue about it: a
blind policy carrying a FLAME-style EWMA offset
($\mathrm{off} \leftarrow (1{-}\alpha)\,\mathrm{off} + \alpha\,(r_t -
\hat{T}(f_t))$, gain $\alpha{=}0.1$, offset initialized to zero), re-picking
the clock every cycle with the same prediction-plus-margin rule as the
static policies and \emph{no} hysteresis --- switches actuate immediately in
the post-completion slack --- run live on the NX (three repeats per
deadline) and replayed over the logged per-cycle traces. In replay the
escape time is insensitive to the gain ($5$--$73$ cycles across
$\alpha \in \{0.02, 0.1, 0.3\}$); a hysteresis band would damp the
knife-edge hunting below at the cost of slower reaction, a tuning we do not
explore. For
this level-shift workload the calibrator escapes the infeasible pick
quickly --- within $1$--$8$ cycles at $D{=}4.8$--$5.0$\,ms, paying $0$--$6$
missed deadlines in the transient --- so the blind failure is a bounded
transient \emph{when} a well-tuned feedback loop runs. Three costs remain,
and they are what the EMC-aware table buys out. First, at the tightest
deadline ($D{=}4.6$\,ms) the live controller never settles: it keeps
crossing the knife-edge boundary between the top two clocks ($13$--$15$
switches per $2{,}000$-cycle run, the last beyond cycle $1{,}500$) --- an
oscillation the stationary replay does not predict, and each switch is an
actuation event. Second, a level offset structurally cannot repair the
memory-clock-dependent \emph{slope}: the memory-bound worst case stays
$20$--$29\%$ (Table~\ref{tab:offset}) no matter how long it converges.
Third, for LLM decode the level itself is non-stationary --- KV-cache
growth drifts per-token latency upward within a decode window (measured in
the Qwen deployment below) --- so a mean-tracking loop chases a moving target
near the cliff. The EMC-aware policy pays none of these: right operating
point from the first release, zero transient, zero hunting, no gain to
tune.

\subsection{Margin provenance and sensitivity}
The safety multiplier $r{=}1.035$ was fixed before any policy evaluation
in this study and held constant across every policy and contract. Its
magnitude upper-bounds the largest solo locked-clock p99/p50 dispersion
observed at the profiling point (${\le}1.9\%$ across all cells) with
roughly $2\times$ headroom. Replaying every evaluated contract of the
main text at $r \in \{1.02, 1.035, 1.05\}$ over the deployed traces:
all vision co-tenancy picks --- point-estimate and guarded, both
deadlines --- are unchanged across the range; the horizon cap moves
($H^{*}(D{=}28) = 3{,}066 / 2{,}147 / 1{,}254$ tokens), so $r{=}1.05$
conservatively refuses the measured-feasible $H{=}2{,}000$ contract that
$r{\le}1.035$ accepts ($0.52\%$ measured); and the joint-stress first
accepted deadline moves from $D{=}44$\,ms ($0.11\%$ measured) to
$D{=}45$\,ms ($0.08\%$) at $r{=}1.05$. No value of $r$ in the range
produces a false admission; a larger $r$ only trades utilization for
conservatism.

\begin{table}[ht]
  \centering
  \caption{Safety-multiplier sensitivity: every evaluated contract
  replayed at three values of $r$ over the deployed traces.}
  \footnotesize
  \begin{tabular}{lrrrr}
    \toprule
    $r$ & co-tenancy picks changed & $H^{*}(D{=}28)$ & joint first-accept $D$ & false adm. \\
    \midrule
    1.02  & 0 & $3{,}066$ ($H{=}2$k accepted) & 44\,ms ($0.11\%$) & 0 \\
    1.035 & 0 & $2{,}147$ ($H{=}2$k accepted) & 44\,ms ($0.11\%$) & 0 \\
    1.05  & 0 & $1{,}254$ ($H{=}2$k declined) & 45\,ms ($0.08\%$) & 0 \\
    \bottomrule
  \end{tabular}
\end{table}

\subsection{Probe protocol}
The occupancy probe is measured once per stable deployment condition
(one $200$-cycle vision probe, one $200$-token decode probe); the
independent deployment runs reported in the main text validate the
selected contract's execution under that condition, not repeated
probing. As one indication of probe stability, the vision probe's
$M{=}3.52$\,ms agrees to $0.01$\,ms with the value implied by the
independent co-tenancy characterization traces collected on a different
day.

\subsection{End-to-end pipeline validation}
Sixteen fresh-probe trials, each with a freshly launched co-runner, a
fresh probe, a fresh guarded decision, and a disjoint production
contract. Slack is the admitted clock's distance from its feasibility
boundary in units of the probe margin.

\begin{table}[ht]
  \centering\footnotesize
  \caption{Fresh-probe admission-loop trials (vision: MobileNetV2,
  $D{=}12$\,ms, guarded rule; decode: Qwen+ViT joint, $H{=}2{,}000$).}
  \begin{tabular}{rlrrrl}
    \toprule
    trial & arm & $M_{\mathrm{g}}$ & decision & slack & miss\,\% \\
    \midrule
    1 & vision & 4.25\,ms & 918\,MHz & $714\,\mu$s & 0.00 \\
    2 & vision & 4.22\,ms & 918\,MHz & $748\,\mu$s & 0.00 \\
    3 & vision & 3.88\,ms & 816\,MHz & $231\,\mu$s & \textbf{3.00} \\
    4 & vision & 4.33\,ms & 918\,MHz & $634\,\mu$s & 1.40 \\
    5 & vision & 4.34\,ms & 918\,MHz & $626\,\mu$s & 0.70 \\
    6 & vision & 3.98\,ms & 816\,MHz & $138\,\mu$s & \textbf{3.10} \\
    7 & vision & 4.42\,ms & 918\,MHz & $548\,\mu$s & 0.00 \\
    8 & vision & 4.55\,ms & 918\,MHz & $415\,\mu$s & 1.40 \\
    \addlinespace
    1 & decode & 15.9\,ms & $D{=}45$ & --- & 0.10 \\
    2 & decode & 17.2\,ms & $D{=}46$ & --- & 0.00 \\
    3 & decode & 17.2\,ms & $D{=}46$ & --- & 0.00 \\
    4 & decode & 17.5\,ms & $D{=}46$ & --- & 0.00 \\
    5 & decode & 17.1\,ms & $D{=}46$ & --- & 0.15 \\
    6 & decode & 16.9\,ms & $D{=}45$ & --- & 0.10 \\
    7 & decode & 17.0\,ms & $D{=}46$ & --- & 0.05 \\
    8 & decode & 16.8\,ms & $D{=}45$ & --- & 0.10 \\
    \bottomrule
  \end{tabular}
\end{table}

Both budget violations (trials 3 and 6) admitted the knife-edge
816\,MHz cell with less than $240\,\mu$s of feasibility slack ---
inside the probe margin's cross-trial dispersion (s.d.\
$199\,\mu$s, range $667\,\mu$s over nine probes) --- while every
passing vision admission carried at least $415\,\mu$s; this
motivates the dispersion band of the main text. In the ten held-out
banded trials (fresh co-runner launch each; band $h{=}400\,\mu$s
fixed pre-validation) the band fired on four probes that would
otherwise have admitted 816\,MHz; every trial selected 918\,MHz,
and the contracts measured $0.0$--$2.9\%$ per launch (aggregate
$1.19\%$, launch-level upper bound $1.70\%$; two of ten windows
above $\tau$), with probe margin and launch outcome uncorrelated
($+0.08$) --- the residual spread is co-tenant launch/phase
realization, not probe error. The banded policy's validation set is the ten held-out launches
($1.19\%$, launch-level upper bound $1.70\%$); pooling all sixteen
same-clock 918\,MHz launches, including the six selected by the
unbanded rule, gives a descriptive $0.96\%$ [$1.34\%$] that we do
not use to validate the banded policy. One bin above the banded pick
(1020\,MHz, eight further independent launches), every launch
measures $0.00\%$ at $D{=}12$\,ms: per-launch compliance is
purchasable with one step of headroom. The decode arm's fresh probes
returned uniformly larger margins and all eight accepted contracts
measured ${\le}0.15\%$.

\subsection{Band-width sensitivity}
Replaying the banded decision over all nineteen collected probes
(reference, round-1, and held-out) at candidate band widths: with
$h{=}0$ six probes admit 816\,MHz; $h{=}1\sigma$ leaves the two
observed failures; any $h$ between $1.5\sigma$ and $2\sigma$
selects 918\,MHz for all nineteen; $h{=}2.5\sigma$ promotes the
four largest-margin probes to 1020\,MHz. The adopted $2\sigma$ is
therefore not knife-edge minimal, and the held-out picks are identical
for any band in the $1.5$--$2\sigma$ range.

\begin{table}[ht]
  \centering\footnotesize
  \caption{Band-width sensitivity: pick distribution over the nineteen
  collected probes.}
  \begin{tabular}{lrrr}
    \toprule
    $h$ & 816\,MHz & 918\,MHz & 1020\,MHz \\
    \midrule
    $0$          & 6 & 13 & 0 \\
    $1\sigma$   & 2 & 17 & 0 \\
    $1.5\sigma$ & 0 & 19 & 0 \\
    $2\sigma$ (adopted) & 0 & 19 & 0 \\
    $2.5\sigma$ & 0 & 15 & 4 \\
    \bottomrule
  \end{tabular}
\end{table}

\section{RQ1: The Memory-Clock Axis}\label{sec:rq1}

\begin{figure}[t]\centering\includegraphics[width=\linewidth]{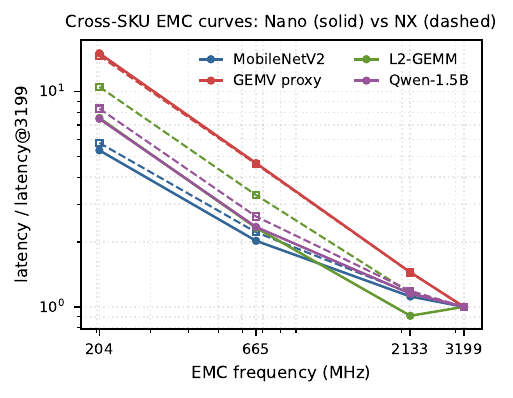}
\caption{Cross-SKU EMC curves (median latency, normalized to 3199\,MHz): Orin Nano (solid) and Orin NX (dashed). The EMC penalty and the universal low-clock collapse replicate on both SKUs; the L2-resident GEMM's non-monotonic inversion (Nano latency \emph{rising} from 2133 to 3199\,MHz) is absent on the NX.}\label{fig:rq1crossku}\end{figure}

\begin{figure}[t]
  \centering
  \includegraphics[width=\linewidth]{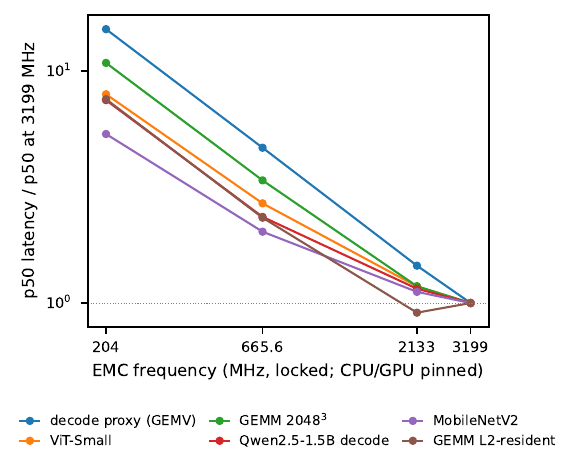}
  \caption{Median latency vs.\ locked EMC frequency (normalized to
  3199\,MHz; CPU/GPU pinned; 1k iterations/cell). Every workload, including
  the near-zero-DRAM L2-resident GEMM, slows several-fold at low memory
  clocks; the
  L2-resident GEMM is \emph{faster} at 2133 than at 3199\,MHz.}
  \label{fig:rq1curves}
\end{figure}

\begin{figure}[t]
  \centering
  \includegraphics[width=\linewidth]{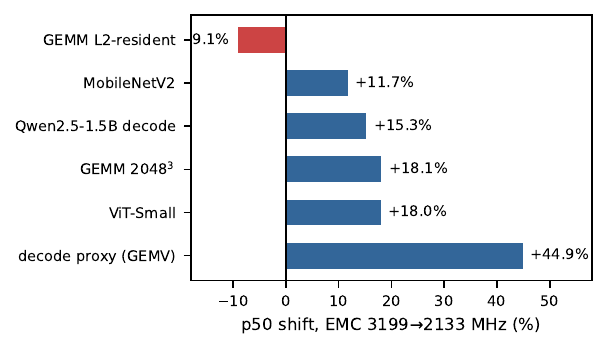}
  \caption{Workload-dependence of the EMC effect in the realistic upper
  range (3199$\to$2133\,MHz): $+45\%$ for bandwidth-bound decode to
  $-9.1\%$ (faster at the lower clock) for L2-resident compute. A
  frequency-independent constant cannot represent this term.}
  \label{fig:rq1spectrum}
\end{figure}

Figure~\ref{fig:rq1curves} shows median latency for all six workloads at the
four lockable EMC points (CPU/GPU pinned, each EMC point locked through the BPMP
interface with readback verification, 1k iterations per cell;
main text \S III). At the bottom of the range the effect is universal:
stepping 665.6 to 204\,MHz raises median latency by $+163\%$ (MobileNetV2) to
$+224\%$ (GEMV decode proxy, L2-resident GEMM) --- roughly a $3\times$ inflation
on this single segment for every workload, regardless of arithmetic intensity.

The most instructive curve is the one that should have been flat. The
L2-resident GEMM serves its weights from L2 with near-zero DRAM streaming, yet
at 204\,MHz it runs $8.3\times$ slower than at its fastest point with the GPU
clock held at 1016--1018\,MHz throughout --- neither a GPU-clock excursion nor a
bandwidth shortfall. The EMC domain gates resources beyond DRAM bandwidth, and
the assumption that a compute-bound kernel is immune to the memory clock is
false on this SoC.

The 204\,MHz floor is a point a deadline-driven governor would rarely select;
the decision-relevant range is the upper pair, $2133\to3199$\,MHz, where the
effect becomes strongly workload-dependent (Fig.~\ref{fig:rq1spectrum}):
lowering 3199 to 2133\,MHz shifts the median by $+45\%$ for the GEMV decode
proxy, $+18\%$ for ViT-Small and the GEMM proxy, $+12\%$ for MobileNetV2, and
$+15.3\%$ for the real SLM (Qwen2.5-1.5B) --- a spectrum of $+11\%$ to $+48\%$
across our campaign and pilot runs, all well above noise (at 3199\,MHz the
p99/p50 ratio is at most 1.019 in every cell). The SLM sits below its GEMV
anchor because Q4\_K\_M dequantization adds per-weight compute that dilutes the
memory-boundedness of batch-1 decode~\cite{decodemembound,decodemembound2}.

\subsection{The penalty replicates across SKUs}\label{sec:rq1crosssku}

To test whether the EMC axis is a property of this SoC family or of one unit,
we repeated the upper-range sweep on a second SKU, an Orin NX 16\,GB on a
third-party carrier under L4T R36.4.3. The boards differ in GPU ceiling (NX
1173\,MHz vs.\ Nano 1020\,MHz), L4T release, and carrier --- confounds we own
rather than control --- but share the same lockable EMC frequency points:
both expose $\{204, 665.6, 2133, 3199\}$\,MHz, making the memory-clock axis
the matched, verified cross-SKU comparison, where agreement despite differing GPU ceilings
and software stacks is the more telling.

On the EMC axis the central findings replicate. The $2133\to3199$ penalty is
of the same magnitude and sign on the NX --- MobileNetV2 $+15.8\%$, the GEMV
proxy $+44.5\%$, the GEMM proxy $+16.1\%$, and the SLM $+18.7\%$ --- against the
Nano's $+11.7\%/+44.8\%/{-}9.1\%/+15.3\%$. These magnitudes are
session-stable: three independent same-day sessions (thermal-gated,
5-minute cooldowns) reproduce every upper-pair penalty within a
${\le}1$\,pp band on the Nano ($3$--$4$\,pp for the bandwidth-bound proxy)
and ${\le}2$\,pp on the NX for five of six workloads --- the exception is
the NX MobileNetV2 response cell, whose knife-edge spans $7$\,pp
($+11.2$ to $+18.5\%$) across sessions, consistent with its run-variable
tail in main text \S IV. The inversion reproduces in all three
sessions ($-9.4$ to $-10.1\%$). The ranking transfers intact: the
bandwidth-bound GEMV proxy is the most EMC-sensitive workload on both boards by
a wide margin, the dense networks an order of magnitude below. The low-clock
collapse transfers too: stepping $665.6\to204$\,MHz inflates the median by
roughly $3\times$ for every workload on the NX as on the Nano, and the full
$204$-vs-$3199$ span reaches $5$--$15\times$ on the Nano and $6$--$15\times$ on the NX. The EMC is a missing axis on
both members of the family, not an idiosyncrasy of one unit.

Table~\ref{tab:catalogue} consolidates the axis in one place: median latency
at every lockable EMC point, per workload, per SKU, per runtime --- the
per-lockable-point table our prescription says a governor should profile
(\S\ref{sec:rq1cost}), instantiated for the six workloads of this study.
The axis also survives --- indeed sharpens under --- the most common
throughput optimization: batching. Sweeping MobileNetV2 at batch
$\{1,4,8\}$ on the Nano, the per-image $2133{\to}3199$\,MHz penalty
\emph{grows} from $+11.9\%$ to $+18.0\%$ to $+20.8\%$, and the 204\,MHz
collapse from $5.4{\times}$ to $9.7{\times}$: batching amortizes kernel
launches but scales the per-batch activation traffic, moving the workload
\emph{down} the roofline~\cite{roofline} toward the bandwidth ridge, so a batched deployment
is more EMC-sensitive, not less.

\begin{table*}[t]
  \centering
  \caption{The EMC axis, consolidated: median latency (ms) at each lockable
  EMC point, with the ratio to the 3199\,MHz point in parentheses. This is
  the per-lockable-point profile the paper prescribes, instantiated for our
  workloads (Nano: GPU pinned 1020\,MHz; NX: 1173\,MHz; 1{,}000-iteration
  cells, CPU pinned; Qwen rows are ms/token). Bold marks the one
  non-monotone cell (the ONNX Runtime L2-resident inversion). NX ViT was
  swept only over the deployment pair. Under TensorRT fp16 the 204\,MHz
  floor is catastrophic for the bandwidth-bound kernels because the fused
  engines saturate memory earlier ($5.3$/$11.8$\,s --- included to warn
  against profiling below the deployment range).}
  \label{tab:catalogue}
  \footnotesize
  \resizebox{\textwidth}{!}{%
  \begin{tabular}{@{}l rrrr rrrr rrrr@{}}
    \toprule
    & \multicolumn{4}{c}{Nano, ONNX Runtime} & \multicolumn{4}{c}{Nano, TensorRT fp16} & \multicolumn{4}{c}{NX, ONNX Runtime} \\
    \cmidrule(lr){2-5}\cmidrule(lr){6-9}\cmidrule(lr){10-13}
    workload & 204 & 665.6 & 2133 & 3199 & 204 & 665.6 & 2133 & 3199 & 204 & 665.6 & 2133 & 3199 \\
    \midrule
    MobileNetV2 & 22.5 (5.3) & 8.5 (2.0) & 4.7 (1.1) & 4.2 & 5.0 (3.3) & 2.4 (1.6) & 1.7 (1.1) & 1.5 & 23.7 (5.8) & 9.1 (2.2) & 4.8 (1.2) & 4.1 \\
    ViT-Small & 90.4 (7.9) & 30.7 (2.7) & 13.5 (1.2) & 11.4 & 16.8 (5.0) & 6.6 (1.9) & 3.6 (1.1) & 3.4 & -- & -- & 12.9 (1.2) & 10.5 \\
    GEMV $12{\times}4096^2$ & 68.6 (15.1) & 21.2 (4.7) & 6.6 (1.4) & 4.6 & 5272 (929) & 21.1 (3.7) & 6.9 (1.2) & 5.7 & 71.1 (14.7) & 22.3 (4.6) & 7.0 (1.4) & 4.8 \\
    GEMM $4{\times}2048^3$ & 118.7 (10.8) & 37.1 (3.4) & 13.0 (1.2) & 11.0 & -- & -- & -- & -- & -- & -- & -- & -- \\
    GEMM $64{\times}1024^3$ & 140.4 (7.5) & 43.4 (2.3) & \textbf{16.9 (0.91)} & 18.6 & 11844 (698) & 360 (21) & 18.3 (1.1) & 17.0 & 146.0 (10.5) & 45.9 (3.3) & 16.1 (1.2) & 13.9 \\
    Qwen2.5-1.5B (ms/tok) & 181.4 (7.5) & 57.0 (2.3) & 28.0 (1.2) & 24.3 & -- & -- & -- & -- & 183.4 (8.3) & 57.9 (2.6) & 26.1 (1.2) & 22.0 \\
    \bottomrule
  \end{tabular}}
\end{table*}

One finding does \emph{not} replicate, and the discrepancy is itself the
result: the penalty, the ranking, and the collapse are platform-level, but the
inversion is configuration-specific. On the Nano the L2-resident GEMM is
$9.1\%$ \emph{faster} at 2133 than at 3199\,MHz (16.93 vs.\ 18.63\,ms median),
reproduced three times independently --- the initial pilot ($-9.9\%$), a cold
re-run started at 48\,$^\circ$C with the GPU clock verified at 1016--1018\,MHz,
and the full campaign ($-9.1\%$) --- so neither a thermal artifact nor a
GPU-clock excursion. It is not a law of the platform: it is specific to the top
GPU clock --- on an unconfounded eight-point GPU sweep at locked EMC (same
profile, only the GPU clock varies), the ordering is normal at every clock up
to $918$\,MHz ($-1.3$ to $-4.4\%$) and flips to $+11.8\%$ only at
$1020$\,MHz, with a tail precursor already visible at $918$\,MHz (per-kernel
slow-mode occupancy ${\sim}25\%$, \S\ref{sec:mechanism}); specific to the SKU (on the Orin NX, identical EMC
points and runtime, the GEMM shows the normal $+16.1\%$ ordering, the first
direct cross-SKU evidence, \S\ref{sec:rq1crosssku}); and specific to the
runtime (under a TensorRT fused engine the inversion is gone and monotonicity
restored, \S\ref{sec:runtime}). It is therefore an \emph{interaction}
among GPU operating point, SKU, and runtime, not a property of the EMC axis
alone --- consistent with \S\ref{sec:rq1cost}, where the GPU-frequency model's
error on the L2-resident compute proxy shrinks at lower GPU clocks. The
evidence is consistent with a sporadic, EMC-dependent memory-fabric slow
state, analyzed in \S\ref{sec:mechanism} (the responsible event remains
unidentified); on the
energy axis the lower-clock point is Pareto-dominant (\S\ref{sec:energy}).
Rather than a universal hazard, it is an \emph{existence proof} that latency
monotonicity in the memory clock \emph{can} fail in a deployed configuration ---
here the conjunction of the Nano SKU, the top GPU clock, and ONNX Runtime's
cuBLAS GEMM.

Three consequences follow for frequency-aware estimators. First, the constant
$b$ that absorbs ``memory transfer delays''~\cite{flame2026} is in reality
$b(f_{\mathrm{emc}}, w)$: a term moving the median by $+45\%$ for one workload
and $-9.1\%$ for another, with CPU and GPU clocks fixed, is neither frequency-
nor workload-independent. Second, frequency-search procedures that assume
latency is non-increasing in clock are unsafe across the inversion --- a
governor raising the memory clock to buy deadline margin can lose it instead
--- so safe search needs measured curves, not assumed shapes. Third, the fix is
bounded: the EMC exposes four lockable points, so adding the axis multiplies
profiling cost by four, not by a continuum.

\subsection{What the omission costs an estimator}\label{sec:rq1cost}

A natural defense of the CPU$\times$GPU formulation is that if the EMC is
fixed during both profiling and deployment, $b$ is a constant of that
configuration. The EMC is indeed not governed dynamically here: under the
stock governor with CPU and GPU pinned, we observe zero EMC transitions across
idle, CNN, GEMV, and SLM workloads --- the clock sits at 2133\,MHz throughout,
even at 45\% EMC utilization under SLM decode in the uncapped power mode,
counterintuitively \emph{below} the 3199\,MHz the capped 25\,W profile pins
(Table~\ref{tab:powermode}). But \emph{which} constant belongs to the platform,
not the model: the same binary under the 15\,W and 25\,W nvpmodel profiles sees
2133 and 3199\,MHz respectively, and a CPU$\times$GPU estimator has no input
distinguishing the two deployments --- the pair our
$2133{\leftrightarrow}3199$ comparisons quantify.

\begin{table}[t]
  \centering
  \caption{EMC behavior under the stock configuration. ``Observed''
  is the EMC clock during 30--90\,s runs with CPU/GPU pinned, sampled at
  10\,Hz; no transition was observed in any trace. The operative point
  differs across profiles (2133 under MAXN\_SUPER, 3199 under 25\,W),
  unrelated to workload demand.}
  \label{tab:powermode}
  \footnotesize
  \resizebox{\columnwidth}{!}{%
  \begin{tabular}{@{}llll@{}}
    \toprule
    nvpmodel profile & EMC config & observed under load & trans.\\
    \midrule
    15\,W & max 2133\,MHz & 2133\,MHz (all 4 traces) & 0\\
    25\,W & max 3199\,MHz & 3199\,MHz in all four traces & 0\\
    MAXN\_SUPER & uncapped & 2133\,MHz (all 4 traces) & 0\\
    \bottomrule
  \end{tabular}}
\end{table}

To measure the cost directly, we fit $T(f_{\mathrm{gpu}}) =
k/f_{\mathrm{gpu}} + b$ per workload on an eight-point GPU devfreq sweep
(306--1020\,MHz, 300 iterations per cell, CPU pinned) at EMC 3199\,MHz, then
evaluated it at the other lockable points (Table~\ref{tab:estimator}). This is
deliberately the \emph{GPU-frequency slice} of the model
family~\cite{flame2026}: even the simplest slice degrades off-profile, and a
richer CPU$\times$GPU \emph{static} model still has no frequency input
distinguishing the two deployments --- only a runtime feedback path (online
drift calibration~\cite{flame2026}) could partially compensate, which we
quantified in the deployment of main text \S IV. In scope the form is excellent --- median
residual 0.3--2.5\%. One lockable point away, at 2133\,MHz, the median error
grows to 3.4--5.3\% with maxima of 12.8--32.2\%; at 665.6\,MHz the median is
38--69\%. The signed error is negative for every workload: the estimator
\emph{under}estimates latency, the direction that converts model error into
deadline misses.

Repairing the model is not as simple as adding a term. A parametric extension
$T = k/f_{\mathrm{gpu}} + m/f_{\mathrm{emc}} + b$, fitted on the 3199 and
665.6\,MHz sweeps and interpolated to the held-out 2133\,MHz point, is
\emph{worse} than the unrepaired model for three of four workloads
(Table~\ref{tab:estimator}): the EMC response is not $1/f$-shaped, and for the
L2-resident GEMM not even monotonic. The closest prior model fits latency as a
monotone power-law sum
$T = A\,f_{\mathrm{gpu}}^{-\alpha} + B\,f_{\mathrm{emc}}^{-\beta}$~\cite{jointmemfreq2025}
(we add a constant offset $C$ when fitting); on our Nano $\text{GPU}\times\text{EMC}$ grid
(20 cells) it fits the monotone workloads well (MobileNetV2 $1.2\%$ in-sample,
$2.7\%$ at a held-out 2133\,MHz cell; ViT-Small $2.7\%/6.4\%$) but generalizes
poorly for the memory-bound kernels (GEMV proxy $16.7\%$ median, $31\%$ max;
L2-resident GEMM $14.8\%/22\%$, against the $2$--$9\%$ a two-cell tabulation
achieves), and for the L2-resident GEMM it gets the inversion \emph{sign} wrong,
predicting $+2.67$\,ms where the measured difference is $-1.70$\,ms --- a
monotone power-law sum structurally cannot represent a curve that turns around.
Tabulation works where it fails: refitting $(k, b)$ from two cells at the target
EMC point bounds the median error to 2.9--4.6\% for three of four workloads, and
the repair holds at 665.6\,MHz (median $2.1\%$/$1.8\%$/$0.0\%$ on the
non-training cells for MobileNetV2/ViT/proxy, $8.3\%$ for the L2-resident
GEMM --- the proxy is fully bandwidth-saturated there, so the two-cell line
fits its nearly GPU-independent latency). The bandwidth-bound
GEMV proxy resists even this (12.7\%/20.9\%): its GPU-frequency dependence itself
changes with the memory clock --- the terms interact --- so it needs the full
per-point sweep. The prescription is also robust to \emph{which} two cells
are profiled: enumerating all $28$ two-cell pairs, the repair holds a
${\le}10\%$ worst-case for $26/28$ pairs on MobileNetV2, $23/28$ on the
L2-resident GEMM, and $15/28$ on ViT --- the failures are adjacent-cell
pairs with too short a frequency lever arm, and the endpoint pair we use is
within $1$--$3$\,pp of the best pair --- while the GEMV proxy fails for all
$28$, confirming that no two-cell refit substitutes for its full sweep. The practical prescription is a per-lockable-point table,
$T = k(f_{\mathrm{emc}})/f_{\mathrm{gpu}} + b(f_{\mathrm{emc}}, w)$: at four
lockable points, a $4\times$ profiling multiplier, and the same profiling pass
yields the tail margins of \S\ref{sec:rq2}.

\begin{table}[t]
  \centering
  \caption{Median/max relative p50 error (\%) on the held-out EMC
  2133\,MHz sweep. A: $k/f_{\mathrm{gpu}}{+}b$ fitted at 3199\,MHz;
  B: ${+}m/f_{\mathrm{emc}}$ fitted at $\{3199, 665.6\}$;
  C: $(k,b)$ refitted from the two endpoint cells at 2133\,MHz (GPU 306 and
  1020\,MHz) and scored on the six non-training cells only.}
  \label{tab:estimator}
  \resizebox{\columnwidth}{!}{%
  \begin{tabular}{lrrr}
    \toprule
    workload & A: GPU-only & B: ${+}m/f_{\mathrm{emc}}$ & C: 2-cell refit \\
    \midrule
    MobileNetV2      & 5.3 / 12.8 & 5.4 / 7.1   & 2.9 / 3.4 \\
    ViT-Small        & 5.1 / 17.6 & 10.5 / 14.1 & 4.6 / 6.7 \\
    decode proxy     & 3.4 / 32.2 & 19.9 / 30.9 & 12.7 / 20.9 \\
    GEMM L2-resident & 1.5 / 2.8  & 17.8 / 34.5 & 2.9 / 4.8 \\
    \bottomrule
  \end{tabular}}
\end{table}

\section{Energy and Pareto-Dominated Operating Points}\label{sec:energy}

\begin{figure}[t]\centering\includegraphics[width=\linewidth]{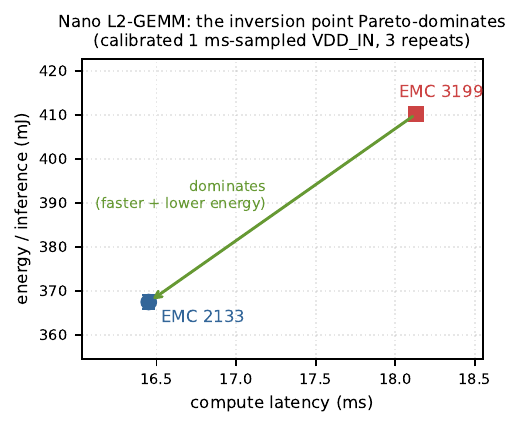}
\caption{The L2-resident GEMM on the latency--energy plane (calibrated $1$\,ms-sampled \texttt{VDD\_IN} integration; EMC point labelled). On the Nano, 2133\,MHz is both faster \emph{and} lower-energy than 3199\,MHz ($368$ vs.\ $410$\,mJ/inf) --- the inversion point Pareto-dominates the maximum. On the NX (no inversion) 3199\,MHz is faster but higher-energy, an ordinary trade-off.}\label{fig:energy}\end{figure}

The premise of frequency-aware DVFS is energy: an estimator exists so a
governor can pick the most efficient setting that still meets the deadline.
The cost of the memory clock is, ultimately, an energy cost, and the
inversion of \S\ref{sec:rq1} is most consequential when it is read on the
energy axis. We measure it with calibrated $1$\,ms-sampled \texttt{VDD\_IN}
integration (main text \S III).

The inversion point is not merely faster --- it is \emph{Pareto-dominant}.
On the Orin Nano, run back-to-back, the L2-resident GEMM consumes
$368$\,mJ per inference at 2133\,MHz against $410$\,mJ at 3199\,MHz: the
higher memory clock is both slower (the $-9.1\%$ inversion of
\S\ref{sec:rq1}) and draws marginally more active power ($22.6$ vs.\
$22.3$\,W), so per-inference energy is $11.6\%$ worse
(Fig.~\ref{fig:energy}; $\pm{<}0.5\%$ over three repeats). A governor that
treats frequency as monotone --- raise the clock to buy deadline margin ---
steps from a dominating operating point onto a dominated one, losing latency
\emph{and} energy at once. No CPU$\times$GPU estimator, and no monotone
memory-frequency term, can express an operating point that is strictly
better on both axes than the one above it; the standard search would never
choose it. (An earlier, coarser \texttt{tegrastats}-based estimate put this gap at
$27\%$; that figure conflated idle power at low duty, and the calibrated
$11.6\%$ supersedes it.)

The energy ranking is workload-specific. For the L2-resident kernel 2133 is
strictly energy-optimal (above); for MobileNetV2 the two clocks are
energy-equivalent within $1\%$ ($42.6$ vs.\ $42.8$\,mJ), the top clock's
small latency gain just offsetting its higher power. The memory clock a
power profile pins (\S\ref{sec:rq1cost}) therefore sets not just latency but
the latency--energy trade, and the maximum is, depending on the workload,
either dominated or merely break-even --- never clearly preferable. On the
NX, where the inversion is absent (\S\ref{sec:rq1}), 3199\,MHz is faster but
draws more power --- an ordinary trade-off, not a dominated point ---
exactly the configuration-specificity the cross-SKU comparison predicts.

\section{RQ2: Tails and Bursts under Contention}\label{sec:rq2}

\begin{figure}[t]
  \centering
  \includegraphics[width=\linewidth]{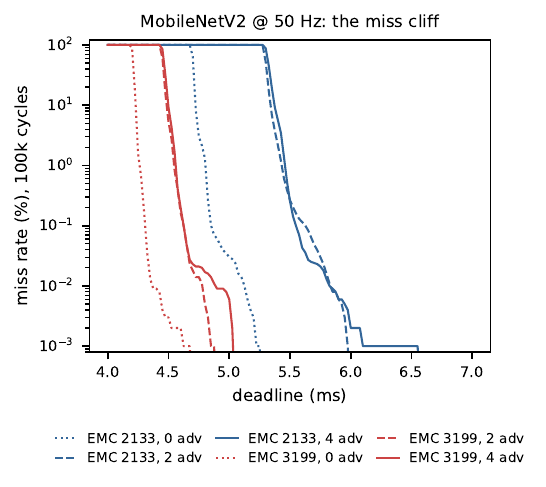}
  \caption{Post-hoc deadline-miss curves (100k cycles/cell). Locked-clock
  distributions are knife-edge: the transition from always-miss to
  never-miss spans ${\sim}1$\,ms. Contention shifts the cliff location
  ($+13\%$ median) more than it fattens the tail.}
  \label{fig:rq2cliffs}
\end{figure}

\begin{figure}[t]
  \centering
  \includegraphics[width=\linewidth]{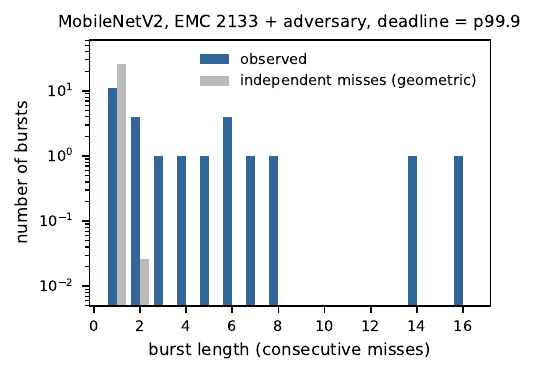}
  \caption{Miss-burst lengths at a p99.9-tight deadline vs.\ the geometric
  distribution that independent misses would produce. Observed bursts reach
  16 consecutive misses; continuation probability is $0.74$ vs.\ $0.001$
  under independence (the worst of four repeated runs; clustering ratio
  $360$--$740\times$ across them, \S\ref{sec:rq2}).}
  \label{fig:rq2bursts}
\end{figure}

\begin{figure}[t]
  \centering
  \includegraphics[width=\linewidth]{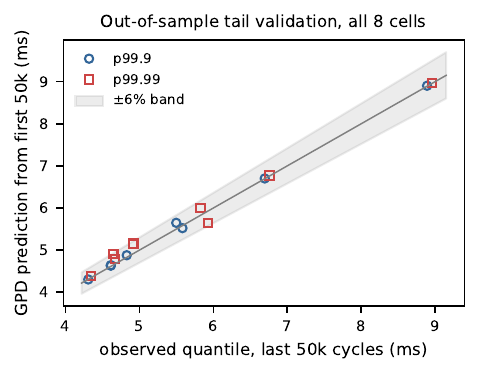}
  \caption{Out-of-sample tail validation: GPD fitted on the first
  50{,}000 cycles of each Part~B cell predicts the last 50{,}000 cycles'
  p99.9 and p99.99 within the $\pm6\%$ band, across all eight cells
  (16 predictions). The in-sample survival illustration is in
  Fig.~\ref{fig:tailfit-appendix}.}
  \label{fig:rq2tailfit}
\end{figure}

We ran 100{,}000 timed cycles in each of eight cells spanning two workloads
(MobileNetV2 and the GEMV decode proxy), two locked EMC frequencies (2133 and
3199\,MHz), and the streaming-write adversary of main text \S III on zero,
two, or four cores (18.9--24.6\,GB/s self-reported). The locked-clock
distributions are knife-edge: p99.99 exceeds the median by ${\le}10\%$
(p99.99/p50 $\le 1.10$) in all eight campaign cells --- across four
independent repeats of the worst cell the ratio spans $1.10$--$1.14$, and
p99.9/p50 stays ${\le}1.11$, so the knife-edge character is run-stable even
where the last decade of the tail is not --- and post-hoc deadline-miss curves
fall off a cliff roughly 1\,ms wide --- with the adversary at 2133\,MHz,
MobileNetV2 misses every cycle at a 5.0\,ms deadline, 0.25--0.26\% at 5.5\,ms,
and at most 0.002\% at 6.0\,ms (Fig.~\ref{fig:rq2cliffs}). Contention mainly
moves this cliff rather than fattening the tail: the two-core adversary shifts
the MobileNetV2 median at 2133\,MHz by $+13\%$ (4.70$\to$5.30\,ms) while
p99.99/p50 grows only from 1.087 to 1.099. The consequence for a deadline
governor is stark: near the cliff a few hundred microseconds of estimation bias
separate 100\% QoS from total failure, so the aggregate miss rate is maximally
sensitive exactly where the estimator must operate.

Aggregate rates also hide \emph{when} misses happen. Anchoring the deadline
post-hoc at each cell's empirical p99.9 --- fixing the aggregate miss rate at
0.1\% (100 misses in 100k cycles) --- we examine the miss sequence. Under
independence the continuation probability (a miss immediately following a miss)
would be 0.001, with geometric bursts of mean length ${\approx}1.001$. Instead
it reaches 0.74 --- 740$\times$ the baseline --- in the worst cell (MobileNetV2,
2133\,MHz, two-core adversary), where the 100 misses collapse into 26 bursts of
mean length 3.85 and maximum 16 (Fig.~\ref{fig:rq2bursts}). The clustering ratio
spans 40--740$\times$ across the six MobileNetV2 cells (single runs per cell)
and 10--50$\times$ for the proxy. Contention is not required: the
\emph{uncontended} MobileNetV2 cell at 2133\,MHz clusters at $540\times$
(continuation probability 0.54), and the contended cells' single-run point
estimates ($340$--$740\times$) sit inside the repeated worst cell's
run-to-run band, so we read no ordering in adversary count from them ---
the adversary moves the cliff more than it creates the bursts.
Replicating the worst cell over four independent 100k-cycle runs (MobileNetV2 /
2133\,MHz / two-core), the clustering ratio is $360$--$740\times$ (mean
$595\times$, s.d.\ $142$): the effect reproduces robustly in magnitude even if
the precise $740\times$ of one run does not. The generalized-Pareto shape, by
contrast, is \emph{not} run-stable ($\xi$ ranges $-0.25$ to $+0.39$), which is
why we rest claims on the clustering ratio and quantiles, not $\xi$. The
signature survives the SKU change: on the Orin NX the same cell clusters
$650$--$720\times$ across three repeats, so strong clustering is a property of
\emph{both} integrated GPUs under contention, not a Nano artifact. What the NX
changes is the stall \emph{magnitude}, not its clustering: its bursts are
heavier-tailed ($\xi=0.6$--$0.8$ across Part~B, single spikes to $12.4$\,ms)
than the Nano's short-tailed ones ($\xi$ near zero --- Gumbel-domain, not bounded support --- spikes to ${\sim}6$\,ms). The
same signature reappears in an independent 1.6M-cycle Nano dataset collected
with a different stressor suite and runner configuration (Appendix~A). Burst \emph{arrivals}
are themselves clustered at 2133\,MHz (inter-burst-interval CV up to 2.6, three
of five cells) but Poisson-consistent at 3199\,MHz (CV 0.88--1.08): the lower
memory clock makes misses both more clustered within bursts and more episodic in
arrival.

The workloads also order opposite to what bandwidth sensitivity predicts.
Despite the largest median EMC sensitivity of any workload ($+45\%$ across
3199$\to$2133\,MHz, \S\ref{sec:rq1}), the GEMV proxy has short tails on every
threshold-robust measure (continuation probability $\le 0.05$, longest burst 3
cycles, extrapolated p99.99 within $3\%$ of median), while MobileNetV2 --- the
\emph{least} EMC-sensitive real model at the median ($+12\%$) --- clusters
hardest and tails widest. So median frequency sensitivity --- what
frequency-aware estimators measure --- failed to predict tail risk in this
two-workload contrast; whether the anticorrelation generalizes is open.

A board-level search returns a bounded negative result: joining per-cycle
timestamps against tegrastats at 500\,ms resolution (burst windows of length
$\ge 2$; 4--57 samples per cell), EMC busy fraction, GPU temperature, and RAM
occupancy are indistinguishable during bursts vs.\ baseline (e.g., 37.0\% vs.\
37.0\% EMC busy, 54.3 vs.\ 54.2\,$^{\circ}$C). Whatever produces the bursts thus
operates below 500\,ms resolution --- pointing to GPU- or driver-level events
rather than thermal or memory-pressure drift --- and we do not identify it.

A GPD fitted to p99 exceedances of each run's first half predicts the held-out
half's p99.9/p99.99 out-of-sample across all eight cells and, at a $0.1\%$
target, calibrates far better than the Gaussian $\mu{+}3\sigma$ margin
(Fig.~\ref{fig:rq2tailfit}); we use it as a marginal quantile heuristic --- the
extrapolated quantile, not $\xi$, is the deliverable, since exceedances
cluster, with threshold stability, bootstrap coverage, declustering, and the
empirical-p99.9 comparison deferred to \S\ref{app:gpd}. Two
prescriptions follow: for offline tail characterization from
sufficiently long traces, quantile margins for deadline governors
should come from EVT fits rather than
$\mu{+}k\sigma$ assumptions --- measurement-based pWCET
practice~\cite{davissurvey} at the governor level --- and QoS reporting,
already moved from means to
percentiles~\cite{percentilereporting,mlperfinference}, should pair the
aggregate rate with the miss-pattern statistics that weakly-hard analysis
consumes~\cite{bernat2001,maggio2020}, e.g.\ the continuation probability
or empirical $\langle m,k\rangle$ violations, since a 0.1\% miss rate
delivered in bursts of sixteen presents a different failure mode than
0.1\% delivered independently. The EVT prescription is scoped to
offline characterization: the deployed admission guard of main text
\S VI is deliberately \emph{maximum}-based, because its short, serially
correlated 200-cycle probe does not support a stable calibrated EVT
margin.

\section{A Shared Stall Signature Consistent with Two Anomalies}\label{sec:mechanism}

\begin{figure}[t]\centering\includegraphics[width=\linewidth]{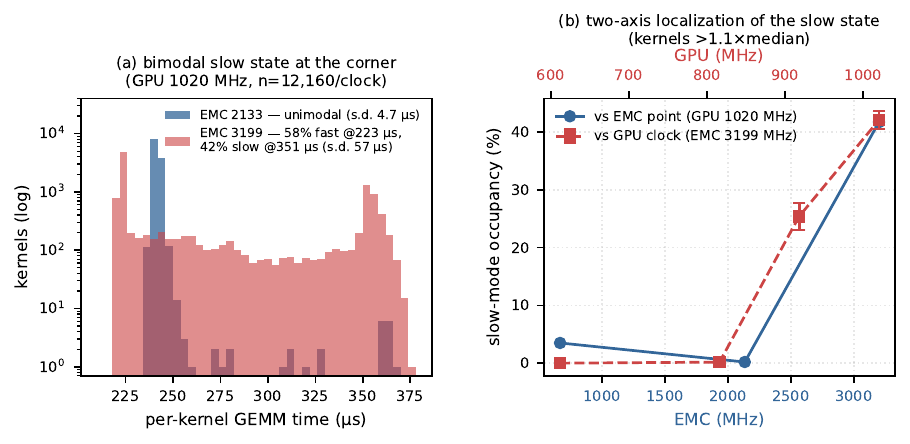}
\caption{The slow state that accounts for the inversion (Nsight Systems, $12{,}160$ kernels per cell, Nano). \textbf{(a)}~At EMC 2133\,MHz per-kernel GEMM times are unimodal (s.d.\ $4.7\,\mu$s); at 3199\,MHz the population splits into a \emph{fast mode} ($58\%$ at $223\,\mu$s --- faster than the 2133\,MHz median, matching the isolated-profiling result) and a \emph{slow mode} ($42\%$ at $351\,\mu$s, $+58\%$). The slow-mode occupancy accounts for the inversion: the per-kernel mean rises $+12.4\%$, against the measured $+11.8\%$ median inversion at the harness. \textbf{(b)}~The slow state is confined to the top of both clock axes: occupancy is $0.2\%$ at EMC 2133 vs.\ $41.7\%$ at 3199 (GPU 1020\,MHz), and ramps $0.2{\to}25{\to}42\%$ across GPU $816{\to}918{\to}1020$\,MHz at EMC 3199 (min--max over three repeats).}\label{fig:mechanism}\end{figure}

Two of our findings resisted explanation in their own right: the
non-monotonic EMC \emph{inversion} of \S\ref{sec:rq1} --- a kernel that runs
faster at a \emph{lower} memory clock --- and the multi-cycle deadline-miss
\emph{bursts} of \S\ref{sec:rq2}. They carry the same statistical signature --- both are consistent with
sporadic, memory-clock-dependent stalls on the integrated GPU that a
per-kernel view cannot see.

\subsection{The inversion is not where it looks}

We instrumented the L2-resident GEMM on the Nano with the GPU clock pinned
at 1020\,MHz, isolating EMC as the only variable, and excluded every local
explanation in turn (Table~\ref{tab:mech-exclusions}). Profiled \emph{in isolation}
with Nsight Compute, each cuBLAS GEMM kernel is $8\%$ \emph{faster} at 3199
than at 2133\,MHz --- fewer long-scoreboard memory stalls, an unchanged
${\sim}83\%$ L2 hit rate --- exactly as expected: run alone, the kernel
likes the faster memory. Preserving cross-kernel cache state
(\texttt{--cache-control none}) does not change this; per-kernel time is flat
across the 64-GEMM weight-sharing sequence and still lower at 3199\,MHz,
excluding both the intra-kernel and the cross-kernel-L2-residency explanations.

The inversion appears only in \emph{real, back-to-back execution} --- and at
kernel resolution it is not a heavy tail but a \emph{bimodal slow state}
(Fig.~\ref{fig:mechanism}a). At 3199\,MHz the per-kernel population splits:
$58\%$ of kernels run in a fast mode at $223\,\mu$s --- \emph{faster} than
the 2133\,MHz median of $241\,\mu$s, exactly as the isolated profile
predicts --- while $42\%$ shift to a slow mode at $351\,\mu$s ($+58\%$). The
occupancy closes the arithmetic: the per-kernel \emph{mean} rises $+12.4\%$
($241.6{\to}271.6\,\mu$s over $12{,}160$ kernels), against the $+11.8\%$
median inversion measured at the harness on an unconfounded eight-point GPU
sweep. A fine-grained trace during the run excludes the obvious culprits:
the GPU clock distribution is identical at the two EMC points
($1013$--$1017$\,MHz, no excursions) and temperature differs by
$1\,^\circ$C, so the slowdown is neither clock throttling nor thermal. What
\emph{does} move is power --- $9$--$15\%$ higher at 3199\,MHz
(\S\ref{sec:energy}).

The slow state is localized on \emph{both} clock axes
(Fig.~\ref{fig:mechanism}b). Sweeping all four lockable EMC points at GPU
1020\,MHz, occupancy is $0.2\%$ at 2133 and $41.7\%$ at 3199\,MHz (the
elevated dispersion at 204--665\,MHz is the memory-starved regime, a
different effect); holding EMC at 3199 and sweeping the GPU clock, occupancy
ramps $0.0{\to}0.2{\to}25{\to}42\%$ across $612{\to}816{\to}918{\to}1020$\,MHz
--- the harness-level median inverts only once occupancy is high enough to
move the mean, which is why the $918$\,MHz cell shows a tail precursor but
normal ordering. In the time domain the slow kernels are not random: they
arrive in runs of \emph{at most three consecutive kernels} (median and
maximum both 3 across $1{,}995$ runs), spread uniformly across inferences
($23$--$32$ slow kernels in every 64-kernel inference) --- consistent with a
size-quantized fabric event of roughly a kernel-triplet's duration
(${\sim}1$\,ms) rather than random arbitration losses or an
inference-scale state. The event is also not the kernel's own code: the NX
runs the \emph{identical} cuBLAS kernel
(\texttt{ampere\_fp16\_s16816gemm\_...\_nn}) with \emph{zero} slow-mode
occupancy at both EMC points, so the slow state is a property of this SKU's
memory fabric at the top $(\mathrm{GPU},\mathrm{EMC})$ clock corner,
consistent with clock-domain-crossing or memory-controller arbitration
behavior specific to that operating region, and we bound the claim to that.
It is likewise implementation- and load-specific: under a TensorRT fused
engine the inversion is gone (\S\ref{sec:runtime}), and a two-core
memory adversary masks it ($+10.2\%$ uncontended vs.\ $-2.4\%$ contended in
the 100k-cycle cells of \S\ref{sec:rq2}) --- consistent with the corner
behavior being masked once the fabric carries concurrent traffic.

\subsection{The miss-bursts carry a matching stall signature}

The deadline-miss bursts of \S\ref{sec:rq2} carry the same signature. In the
worst cell (MobileNetV2, 2133\,MHz, two-core adversary), the cycles in the
extreme tail owe their excess to \emph{compute}, not scheduling: a slow
cycle's compute time exceeds the run mean by $387\,\mu$s on the Nano (and
$621\,\mu$s on the NX), while its release jitter exceeds the mean by only
${\sim}29\,\mu$s. These are GPU stalls, not late releases --- and they
cluster as sporadic stalls would. On the Nano the compute-time series has
lag-1 autocorrelation $0.565$, and a tail cycle follows a tail cycle with
probability $0.74$, $740\times$ the independent rate: the stalls arrive in
persistent runs.

The cross-SKU contrast of \S\ref{sec:rq2} is then a difference in \emph{how
the same signature manifests}, not in whether the stalls occur. Under a two-core
adversary the misses cluster strongly on \emph{both} SKUs --- a
continuation probability ${\sim}600\times$ the independent baseline --- though
the magnitude is run-variable (Nano $360$--$740\times$ across four repeats,
NX comparable), so clustering is not the SKU discriminator. What differs is
the \emph{magnitude} of the individual stall: the NX is consistently
heavier-tailed (generalized-Pareto $\xi{\sim}0.7$ across its cells, single
spikes to $12.4$\,ms) where the Nano is short-tailed ($\xi$ near zero, spikes to
${\sim}6$\,ms). The signature also spans workload classes: extending the 100k-cycle
protocol to ViT-Small and the L2-resident GEMM, ViT clusters as heavily as
MobileNetV2 ($350$--$440\times$ across the four EMC$\times$adversary cells,
maximum burst 20), while the L2-resident kernel barely clusters
($0$--$70\times$, bursts ${\le}6$) even though it is the only workload with
\emph{negative} EMC sensitivity --- median sensitivity and tail risk remain
uncorrelated across five workloads. One measured asymmetry must be stated
rather than smoothed over: the two manifestations express at \emph{opposite}
ends of the EMC axis --- the inversion's slow state opens at the \emph{top}
memory clock (41.7\% occupancy at 3199\,MHz, 0.2\% at 2133), while miss
clustering under contention is strongest at the \emph{lower} deployed clock
(with burst arrivals episodic at 2133 and Poisson-consistent at 3199,
\S\ref{sec:rq2}). A matching compute-side stall signature accompanies both
--- at the top clock corner for the L2-resident kernel it appears as the
bimodal slow state, and at the contended operating point as clustered miss
bursts --- but which clock point expresses it depends on workload,
contention, and SKU. This is the explanatory layer that the aggregate miss
rate and the per-frequency latency model both omit.

\subsection{The mechanism evidence}\label{sec:mech-evidence}
What is striking about the inversion is how little of it survives
isolation. Profiled \emph{in isolation} with Nsight Compute (kernel-replay,
GPU pinned at 1020\,MHz), each cuBLAS GEMM kernel reports the
\emph{opposite} of the deployed ordering --- $8\%$ \emph{faster} at 3199\,MHz
than at 2133\,MHz, L2 hit rate ${\sim}83\%$ at both clocks --- and the
warm-L2 replay is still faster at the higher clock, so the inversion is
neither an intra-kernel nor a cross-kernel L2-residency effect. It appears
only in the \emph{real} back-to-back execution, where an \texttt{nsys} trace
($12{,}160$ kernels per cell) shows the population split into the bimodal
fast/slow modes of Fig.~\ref{fig:mechanism}a: the per-kernel standard
deviation grows from $4.7\,\mu$s to $57\,\mu$s (${\sim}12\times$) and the
per-kernel mean rises $+12.4\%$ --- the inversion. A simultaneous fine-grained GPU-clock, power, and temperature
trace excludes the two obvious board-level confounds: the GPU-clock
distribution is \emph{identical} at 2133 and 3199\,MHz (${\sim}1013$--$1017$\,MHz,
no dips) and die temperature differs by one degree ($51$ vs.\ $52\,^\circ$C),
so the slowdown is neither throttling nor thermal; only supply current rises
($+9$ to $15\%$ at the higher clock). The residual is consistent with a \emph{sporadic}
memory-subsystem/fabric stall at the higher EMC clock --- at constant GPU
clock and temperature the GPU waits for a kernel that barely touches DRAM,
invisible to isolated profiling. We bound our claim to this set of
exclusions; Table~\ref{tab:mech-exclusions} records each hypothesis with the
test that rules it out.

\begin{table}[t]
  \centering
  \footnotesize
  \caption{What the inversion is \emph{not} (Nano, L2-resident GEMM, GPU
  pinned at 1020\,MHz). Each hypothesis is rejected by a separate
  measurement; the residual is consistent with a sporadic memory-fabric stall whose responsible event we do not
  identify here.}
  \label{tab:mech-exclusions}
  \resizebox{\columnwidth}{!}{%
  \begin{tabular}{@{}lll@{}}
    \toprule
    hypothesis & test & verdict \\
    \midrule
    intra-kernel
      & \begin{tabular}[t]{@{}l@{}}isolated \texttt{ncu} replay\\(kernel alone)\end{tabular}
      & \begin{tabular}[t]{@{}l@{}}excluded: $8\%$ \emph{faster}\\at 3199 (L2 hit ${\sim}83\%$)\end{tabular} \\
    \addlinespace
    \begin{tabular}[t]{@{}l@{}}cross-kernel\\L2 residency\end{tabular}
      & \begin{tabular}[t]{@{}l@{}}warm-L2 \texttt{ncu}\\replay\end{tabular}
      & \begin{tabular}[t]{@{}l@{}}excluded: warm-L2 kernel\\still faster at 3199\end{tabular} \\
    \addlinespace
    GPU-clock throttle
      & \begin{tabular}[t]{@{}l@{}}fine GR3D-clock\\trace (real run)\end{tabular}
      & \begin{tabular}[t]{@{}l@{}}excluded: identical dist.\\${\sim}1013$--$1017$\,MHz, no dips\end{tabular} \\
    \addlinespace
    thermal
      & \begin{tabular}[t]{@{}l@{}}die-temp trace\\(real run)\end{tabular}
      & \begin{tabular}[t]{@{}l@{}}excluded: $51$ vs.\ $52\,^\circ$C\\($1\,^\circ$ difference)\end{tabular} \\
    \addlinespace
    \begin{tabular}[t]{@{}l@{}}kernel code /\\cuBLAS selection\end{tabular}
      & \begin{tabular}[t]{@{}l@{}}\texttt{nsys} on the NX\\(same workload)\end{tabular}
      & \begin{tabular}[t]{@{}l@{}}excluded: \emph{identical} kernel,\\zero slow mode at both EMCs\end{tabular} \\
    \addlinespace
    \begin{tabular}[t]{@{}l@{}}random arbitration\\losses\end{tabular}
      & \begin{tabular}[t]{@{}l@{}}timestamped kernel\\events (time domain)\end{tabular}
      & \begin{tabular}[t]{@{}l@{}}excluded: slow kernels arrive\\in runs of ${\le}3$, uniform\\across inferences (quantized)\end{tabular} \\
    \addlinespace
    \begin{tabular}[t]{@{}l@{}}\emph{residual}:\\size-quantized\\fabric slow state\end{tabular}
      & \begin{tabular}[t]{@{}l@{}}\texttt{nsys} real trace,\\4 EMC $\times$ 4 GPU\\clock cells\end{tabular}
      & \begin{tabular}[t]{@{}l@{}}bimodal $58/42\%$ split;\\occupancy $0.2{\to}41.7\%$\\(EMC), $0{\to}42\%$ (GPU);\\mean $+12.4\%$ $\approx$ inversion\end{tabular} \\
    \bottomrule
  \end{tabular}}
\end{table}

\section{RQ3: Frequency Actuation Lag}\label{sec:rq3}

\begin{figure}[t]
  \centering
  \includegraphics[width=\linewidth]{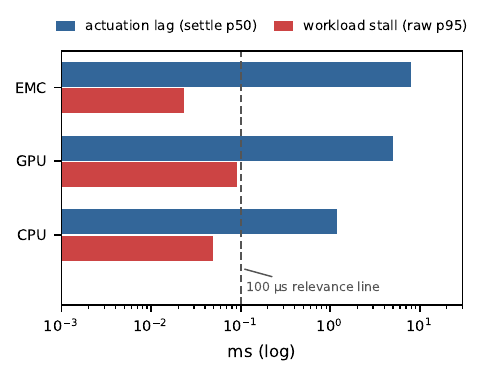}
  \caption{Per-domain transition costs. Workload-observed stalls stay
  below the pre-specified $100\,\mu$s relevance line (raw p95
  ${\le}92\,\mu$s, against matched probe-noise floors of up to
  $84\,\mu$s), but the new frequency takes 1/5/8\,ms (CPU/GPU/EMC) to
  take effect; the EMC's firmware bookkeeping lags a further
  ${\sim}13$\,ms.}
  \label{fig:rq3actuation}
\end{figure}

We pre-specified a demotion criterion for this experiment before running
it: per-domain transition \emph{stalls} would be reported as a headline
result only if the workload-observed stall exceeded $100\,\mu$s at p95
on some domain. The criterion fired against us, and we report the stalls
as the null result they are. We measured two frequency pairs per domain
(CPU $1728\leftrightarrow\{115.2, 729.6\}\,$MHz, GPU
$1020\leftrightarrow\{306, 612\}\,$MHz, EMC
$3199\leftrightarrow\{2133, 665.6\}\,$MHz), 150 (CPU) or 200 (GPU, EMC)
transitions per direction,
using cycle-counted probes that timestamp fixed work quanta on
\texttt{CLOCK\_MONOTONIC} around the frequency write: a dependent-multiply
chain on the CPU, a \texttt{clock64} spin kernel on the GPU, and
DRAM-streaming chunks for the EMC. The worst raw stall at p95 is $49\,\mu$s on the CPU, $92\,\mu$s on the
GPU, and $24\,\mu$s on the EMC (Fig.~\ref{fig:rq3actuation}); matched
no-transition control windows put the probes' own noise floors at
$10$--$84\,\mu$s, so the GPU's $92\,\mu$s sits barely above its
$84\,\mu$s floor. (We report raw quantiles with the noise floor
alongside rather than their difference, which is not itself a quantile.)
All domains fall below the $100\,\mu$s criterion, consistent in magnitude
with the tens of microseconds (roughly 10--70$\,\mu$s) measured for Intel
CPU frequency transitions~\cite{intelfreqtransition}. Rewriting a clock does not stall a
running workload on this board in any way that matters at millisecond
deadlines.

What the same traces do show is \emph{actuation lag}. The probe's
per-quantum rate settles at the new operating point about $1\,$ms after
the write on the CPU (settle p50 $0.99$--$1.40$, p95 ${\le}1.65\,$ms
across pairs and directions), $4.8$--$5.3\,$ms (p95 ${\le}6.3$) on the
GPU, and $7.5$--$8.7\,$ms (p95 ${\le}9.7$) on the EMC --- the lag
distribution is itself tight, so the 1/5/8\,ms framing carries to the
tail. During this window the workload does not
stall---it simply keeps executing at the old frequency. Switching on this
SoC is therefore nearly free in the stall sense, but not in the control
sense studied on discrete GPUs~\cite{gpufreqswitch2025}: a governor's
decision takes roughly 1/5/8\,ms (CPU/GPU/EMC) to become true.

One methodological finding deserves emphasis. The BPMP debugfs
\texttt{rate} readback does not report the target EMC frequency until
$12.9$--$13.8\,$ms (median) after the write---roughly $5\,$ms \emph{after}
the workload demonstrably runs at the new rate. The readback is firmware
bookkeeping, not hardware state; we treat the \texttt{pto\_counter}
readout as ground truth for the EMC rate --- its value (e.g.,
3{,}191{,}887{,}872 against a requested 3{,}199{,}000{,}000) reflects a
measured clock rather than the request table.
The same caution applies on every domain: cpufreq, devfreq, and BPMP
readback files are driver or firmware state, useful for verifying that a
lock took hold, and should never be cited as a measurement of when---or
whether---the hardware transitioned.

\section{Runtime Robustness}\label{sec:runtime}

A reasonable objection is that all this indicts a runtime, not a platform. RQ1
was measured under ONNX Runtime's CUDA provider in fp32~\cite{onnxruntimesw},
which dispatches generic cuBLAS kernels; production edge deployments ship a
fused, offline-scheduled TensorRT engine in fp16~\cite{tensorrt}. If our EMC
sensitivity were an artifact of unfused fp32 dispatch it would evaporate under
the deployed runtime, mooting the missing-axis claim. We therefore added a
TensorRT execution-provider backend (fp16, on-disk engine cache amortizing the
optimization pass across cells; main text \S III) and re-ran the Nano EMC
sweep for the three ONNX workloads that build clean engines plus the two
synthetic proxies, CPU and GPU clocks pinned and each EMC point locked and read
back as before.

The load-bearing finding survives. Stepping the memory clock from 3199 down to
2133\,MHz still inflates the median for every monotone workload under TensorRT
(Table~\ref{tab:trtort-ext}): MobileNetV2 by $+9.9\%$, ViT-Small by $+7.8\%$, the
GEMM proxy by $+20.8\%$. The low-clock collapse is if anything more dramatic: at
204\,MHz the decode proxy runs at 5272\,ms and the L2-resident GEMM at
11844\,ms, the same several-fold floor we report under ONNX Runtime. EMC is a
missing axis under a fused fp16 engine just as under generic fp32 dispatch; the
effect is \emph{runtime-invariant on this platform} in the sense that matters
for an estimator --- fixing CPU and GPU clocks does not fix latency.

What the runtime \emph{does} change is magnitude. Across the three monotone
workloads the $2133\to3199$ penalty under TensorRT is roughly half its ONNX
Runtime value (Table~\ref{tab:trtort-ext}) --- the expected direction, since a
fused engine keeps intermediate activations in registers and shared memory
rather than round-tripping through DRAM, shrinking the latency fraction exposed
to the EMC clock. The axis does not disappear, but a TensorRT deployment sits
at a milder point on it. An estimator calibrated on the wrong runtime would
mis-scale the EMC term even where it correctly includes it; the
per-lockable-point table of \S\ref{sec:rq1cost} must be built under the
deployed runtime.

\begin{table}[t]
  \centering
  \caption{$2133\to3199$\,MHz median penalty (\%; positive = lower clock
  slower) under TensorRT fp16 vs.\ ONNX Runtime CUDA fp32 on the Nano, CPU
  and GPU pinned. The EMC sensitivity persists under both runtimes but is
  roughly halved by the fused engine; the L2-resident GEMM's sign flips ---
  the inversion is present only under ONNX Runtime.}
  \label{tab:trtort-ext}
  \resizebox{\columnwidth}{!}{%
  \begin{tabular}{lrrl}
    \toprule
    workload & TensorRT fp16 & ORT fp32 & note \\
    \midrule
    MobileNetV2      & $+9.9$  & $+11.7$ & monotone, both \\
    ViT-Small        & $+7.8$  & $+18.0$ & monotone, both \\
    GEMV proxy       & $+20.8$ & $+44.8$ & monotone, both \\
    GEMM L2-resident & $+8.1$  & $-9.1$  & inversion: ORT only \\
    \bottomrule
  \end{tabular}}
\end{table}

\subsection{The inversion is a scoped existence proof}\label{sec:inversionscope}

The non-monotonic inversion behaves differently from the rest of the curve.
Under TensorRT the L2-resident GEMM is \emph{faster at the higher memory clock},
in the normal direction: 16.97\,ms at 3199\,MHz vs.\ 18.34\,ms at 2133\,MHz, a
$+8.1\%$ penalty for the lower clock. Under ONNX Runtime the same workload, same
SoC, same GPU and CPU clocks, runs the other way --- $-9.1\%$, faster at 2133
than at 3199\,MHz (\S\ref{sec:rq1}). Changing only the runtime, every clock held
fixed, removes the inversion and restores monotonicity. It is thus not a
property of the EMC axis nor of the SoC alone, but of an \emph{interaction}
between a specific runtime configuration --- ONNX Runtime's cuBLAS kernel for
this GEMM --- and the top of the memory-clock range, on the top GPU clock.

This is consistent with the sporadic-stall signature analyzed
in~\S\ref{sec:mechanism} for this workload: the inversion is not a slower kernel but a heavier
\emph{tail} of kernel execution times appearing at the higher EMC clock, at
unchanged GPU clock and temperature. TensorRT emits a different, fused schedule
for the same GEMM and does not exhibit the signature in our traced
configuration: the tail is absent and ordering is restored. We cannot claim
TensorRT is immune in general, only that the schedule it emits here does not
express the slow state. Note the kernel alone is not the cause either: the NX
runs the \emph{identical} cuBLAS kernel with zero slow-mode occupancy
(\S\ref{sec:mechanism}), so the stall requires \emph{both} this kernel
schedule \emph{and} this SKU's fabric at the top clock corner --- exactly what
one expects if the slow state is an interaction between a kernel's
memory-access pattern and the fabric at high EMC, not a property of the
workload's arithmetic.

We therefore reframe the inversion narrowly. It is not a law of the platform or
a universal hazard of the memory clock; it is an \emph{existence proof} that
latency monotonicity in the memory clock can fail under a deployed runtime,
scoped here to the conjunction (Nano SKU $\times$ top GPU clock $\times$ ONNX
Runtime cuBLAS GEMM). The companion SKU confirms the scoping: on the Orin NX the
same GEMM under ONNX Runtime shows the normal $+16.1\%$ ordering with no
inversion (\S\ref{sec:rq1}), so even the ORT-cuBLAS path does not invert on
every board. The estimation consequence is the point we rest on: a
frequency-search that assumes latency is non-increasing in clock is unsafe,
because we have a reproducible counterexample on real hardware --- and a search
that \emph{cannot} hit the inversion on a given (SKU, runtime, clock) triple can
know so only by measuring the curve, the same per-point tabulation the rest of
this paper argues for. The missing-axis claim is runtime-invariant; the
inversion is the scoped illustration that the axis can be non-monotone, not that
it always is.

\section{Independent burst replication}\label{app:burstrep}
In a separate campaign, the same board ran a 16-cell
sweep (100k cycles each, 1.6M total) with a different runner
configuration and a \texttt{stress-ng}-based stressor suite instead of
the streaming-write adversary. Computing the same continuation statistic
at p99.9-anchored deadlines reproduces the signature: the
memory-pressured cells cluster
($P[\text{miss}\mid\text{miss}] = 0.40$ under the memory stressor, 0.19
and 0.14 under combined memory/VM stressors, 0.11 under an IRQ storm ---
$110$--$400\times$ independence), while baseline, CPU-only, and
cache-only cells sit at $0.00$--$0.07$. Clustering under memory pressure
and near-independence without it replicate across the two campaigns'
different stressors, dates, and runner configurations. Raw data ships in
the repository (\texttt{results/p3*}).

\section{Margin calibration: GPD vs.\ Gaussian}\label{app:gpd}
This appendix expands the margin-calibration result summarized in
\S\ref{sec:rq2}. The extrapolated quantile is stable to the threshold
choice --- fitting the GPD at p98.5, p99, or p99.5 moves the predicted
p99.9 by less than $2\%$ --- even in cells where the shape estimate $\xi$
itself is not threshold-stable (the contended proxy's $\xi$ flips sign
between fitting thresholds), which is why we report the extrapolated
quantile rather than $\xi$. Because exceedances cluster, ordinary-bootstrap
CIs (${\sim}1\%$ half-width) ignore the dependence and cover the empirical
p99.99 in only four of eight cells --- a concrete coverage failure, not a
hypothetical one --- while the empirical p99.99 of a 100k-cycle run is
itself only ${\sim}10$ order statistics. As a dependence check,
runs-declustering (refitting on cluster maxima) moves the predicted p99.99
by at most $5\%$ across gap parameters 1--10 --- most in the most-clustered
cell --- within the out-of-sample error band of the main text. Against a
plain empirical p99.9 from the same profiling window, the GPD is not
uniformly more accurate: the empirical quantile is competitive in most
cells but reaches $4.4\times$ the $0.1\%$ target in its worst cell, where
the GPD margin stays at $1.8\times$ --- the parametric tail buys worst-case
boundedness rather than average accuracy, and the GPD margin itself
achieves $0.016$--$0.20\%$ on held-out data, conservative by up to
$6\times$ in some cells. Finally, the tail is compute-side, not
scheduler-side: release jitter at p99.9 is ${\sim}23$--$37\,\mu$s across
cells, against compute times of 4.3--8.9\,ms, so neither the periodic
harness nor CPU scheduling explains the bursts.

\section{What is ruled out}\label{app:ruledout}
The paper reports three phenomena without a mechanism. Table~\ref{tab:ruledout}
collects the causes excluded by measurement; the remainder is genuinely open.

\begin{table}[ht]
  \centering
  \caption{Causes excluded by a separate measurement for the latency
  inversion (\S\ref{sec:rq1}), the miss-bursts (\S\ref{sec:rq2}), and EMC
  pinning, with the residual mechanisms that remain open.}
  \label{tab:ruledout}
  \footnotesize
  \resizebox{\columnwidth}{!}{%
  \begin{tabular}{@{}lll@{}}
    \toprule
    phenomenon & ruled out (evidence) & still open \\
    \midrule
    \begin{tabular}[t]{@{}l@{}}inversion\\(\S\ref{sec:rq1})\end{tabular}
      & \begin{tabular}[t]{@{}l@{}}thermal (cold re-run, 48$^{\circ}$C)\\
        GPU-clock excursion (held 1016--1018\,MHz)\\
        DRAM-traffic shift (EMC busy\% identical)\\
        single-run noise ($3\times$ reproduced)\end{tabular}
      & \begin{tabular}[t]{@{}l@{}}MC arbitration,\\
        fabric clock ratios\end{tabular} \\
    \addlinespace
    \begin{tabular}[t]{@{}l@{}}miss bursts\\(\S\ref{sec:rq2})\end{tabular}
      & \begin{tabular}[t]{@{}l@{}}harness backlog (max resp.\ $<$ period)\\
        release jitter (p99.9 ${\sim}30\,\mu$s)\\
        RT throttling (disabled; \S\ref{sec:pitfalls})\\
        logging I/O (no I/O in timed loop)\\
        CPU-EP fallback (hard-fail guard)\\
        board state at 500\,ms (EMC\%/temp/RAM flat)\end{tabular}
      & \begin{tabular}[t]{@{}l@{}}sub-500\,ms GPU/\\driver events\end{tabular} \\
    \addlinespace
    \begin{tabular}[t]{@{}l@{}}EMC pinning\\(Table~\ref{tab:powermode})\end{tabular}
      & \begin{tabular}[t]{@{}l@{}}workload demand (45\% busy, no upclock)\end{tabular}
      & \begin{tabular}[t]{@{}l@{}}BPMP/actmon\\policy internals\end{tabular} \\
    \bottomrule
  \end{tabular}}
\end{table}

\section{Supplementary figures}\label{app:figs}

\begin{figure}[ht]
  \centering
  \includegraphics[width=\linewidth]{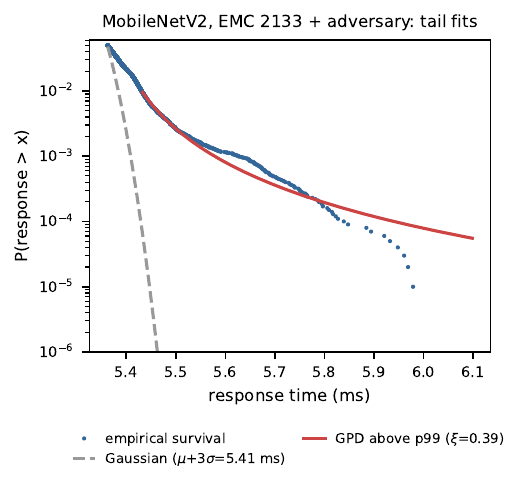}
  \caption{In-sample illustration of the heaviest cell's tail: the
  Gaussian survival collapses where the empirical tail extends; the GPD
  tracks it. The quantitative (out-of-sample) validation is
  Fig.~\ref{fig:rq2tailfit}.}
  \label{fig:tailfit-appendix}
\end{figure}

\begin{figure}[ht]
  \centering
  \includegraphics[width=\linewidth]{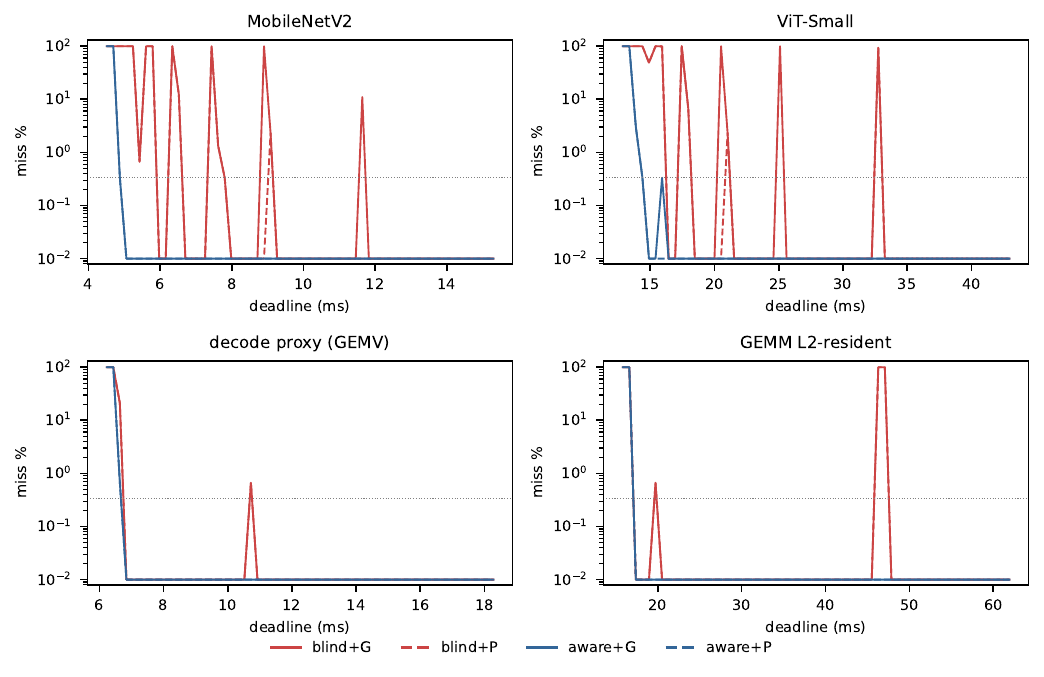}
  \caption{Trace-driven governor \emph{simulation} (complementary to the
  \emph{measured} deployment of main text \S IV, main-text Fig.~2)
  over the full deadline grid for all four workloads: each red spike is the
  EMC-blind policy stepping onto the miss cliff of its selected frequency.}
  \label{fig:govsim-all}
\end{figure}

\bibliographystyle{IEEEtran}
\bibliography{references}

\end{document}